\documentclass[12pt]{article}
\usepackage[utf8]{inputenc}
\usepackage[a4paper, total={6in, 8in}]{geometry}
\usepackage{graphicx} 
\usepackage{amsmath}

\usepackage{amssymb}
\usepackage{array}

\newtheorem{theorem}{Theorem}
\newtheorem{corollary}[theorem]{Corollary}

\newtheorem{lemma}[theorem]{Lemma}
\newtheorem{claim}[theorem]{Claim}
\newtheorem{definition}[theorem]{Definition}
\newtheorem{proposition}[theorem]{Proposition}

\newtheorem{remark}[theorem]{Remark}

\newenvironment{proof}{\noindent\bf{Proof.}\rm}{\hfill$\blacksquare$\bigskip}

\newcommand{\items}{\mathcal{M}} 
\newcommand{\agents}{\mathcal{N}} 

\title{Low communication protocols for fair allocation of indivisible goods}
\author{Uriel Feige\thanks{Weizmann Institute, Israel. {\tt uriel.feige@weizmann.ac.il}}}

\begin{document}

\maketitle

\begin{abstract}
    We study the multi-party randomized communication complexity of computing a fair allocation of $m$ indivisible goods to $n < m$ equally entitled agents. We first consider MMS allocations, allocations that give every agent at least her maximin share. Such allocations are guaranteed to exist for simple classes of valuation functions. We consider the expected number of bits that each agent needs to transmit, on average over all agents. For unit demand valuations, we show that this number is only $O(1)$ (but  $\Theta(\log n)$, if one seeks EF1 allocations instead of MMS allocations), for binary additive valuations we show that it is $\Theta(\log \frac{m}{n})$, and for 2-valued additive valuations we show a lower bound of $\Omega(\frac{m}{n})$. 

    For general additive valuations, MMS allocations need not exist. We consider a notion of {\em approximately proportional} (Aprop) allocations, that approximates proportional allocations in two different senses, being both Prop1 (proportional up to one item), and $\frac{n}{2n-1}$-TPS (getting at least a $\frac{n}{2n-1}$ fraction of the {\em truncated proportional share}, and hence also at least a $\frac{n}{2n-1}$ fraction of the MMS).  We design randomized protocols that output Aprop allocations, in which the expected average number of bits transmitted per agent is $O(\log m)$. For the stronger notion of MXS ({\em minimum EFX share}) we show a lower bound of $\Omega(\frac{m}{n})$.
    
\end{abstract}

\newpage

\section{Introduction}

We consider the multi-party communication complexity of fair allocation of indivisible goods. In our setting, which is the standard setting for such problems, there are $n$ agents and a set $\items$ of $m$ indivisible items that are to be allocated to the agents. An allocation $A$ is a partition of the items into $n$ disjoint sets, $A_1, \ldots, A_n$, where for each $i \in \{1, \ldots, n\}$, set $A_i$ is given to agent $i$. We wish the allocation to be ``fair".

We assume that each agent $i$ has her own private valuation function $v_i$, that maps sets of items to non-negative values. 
We further make the assumption that $v_i$ is completely determined by the value that it gives to individual items, or in our terminology, that $v_i$ is {\em item-based}. Two well known classes of item-based valuation functions are {\em unit demand}, in which $v_i(S) = \max_{e\in S} v_i(e)$ for every set $S \subset \items$, and {\em additive}, in which $v_i(S) = \sum_{e\in S} v_i(e)$.  
(We use $v_i(e)$ is shorthand notation for $v_i(\{e\})$). Note that if each item can have one of $P$ possible known values (e.g., an integer value in the range $[0, P-1]$), then an agent may completely specify her valuation function by communicating $m \log P$ bits. 

In this paper we consider the setting of equal entitlements. That is, all agents should receive the same fairness guarantees. {In our notation, this corresponds to each agent having an entitlement of $\frac{1}{n}$.}

There are several different ways in which fairness can be defined, and we briefly review some of them.

One common framework for defining fairness is by using a {\em share based} approach. Given $\items, n, v_i$, a share function $s$ specifies for each set $S \subset \items$ whether it is {\em acceptable} for agent $i$ to receive the set $S$. An example of a share function is the {\em proportional share}, Prop, for which a set $S$ is acceptable if and only if its value is at least a $\frac{1}{n}$ fraction of the total value. That is, $Prop(v_i, \items, \frac{1}{n}) = \frac{1}{n} v_i(\items)$, and a set $S$ is acceptable for $i$ if and only if $v_i(S) \ge Prop(v_i, \items, \frac{1}{n})$. An allocation is considered acceptable under the share based fairness notion if each agent receives a bundle that is acceptable for her. 

Another common framework for fairness is by using a {\em comparison based} approach. In this approach, an agent is concerned both with the set of items that she herself receives, and with how the remaining items are distributed among the other agents. An example of a comparison based fairness notion is {\em envy freeness}. Under this notion, an agent $i$ views an allocation $A_1, \ldots, A_n$ as acceptable if no other agent receives a bundle that $i$ prefers over $A_i$. That is, $v_i(A_i) \ge v_i(A_j)$ needs to hold for every $j \not= i$. An allocation is {\em envy-free} (EF) if all agents find it acceptable in this sense.

The notions of proportional allocations and envy-free allocations were first introduced in the context of dividing a divisible good among equally entitled agents, also referred to as {\em cake cutting}. When valuations of agents are additive across pieces of the good, 
then every envy-free allocation is also a proportional allocation.  It is known that in this setting, proportional allocations and envy-free allocations always exist. However, there seems to be a big difference in the communication complexity of finding such allocations. There are fairly efficient protocols that find proportional allocations~\cite{EP84}, but known algorithms for finding envy-free allocations are much less efficient~\cite{AM16}.

In the setting of indivisible goods, the one studied in this paper, it may happen that no allocation is proportional or envy-free. For example, this happens whenever $m < n$. Hence it is desirable to instead consider other fairness notions that are {\em feasible}, meaning that there always are allocations acceptable under these notions. Based on the insights gained from the case of a divisible good, one would expect that there will be efficient allocation procedures for share based fairness notions, but perhaps not so for comparison based fairness notions (though of course this depends on the details of the fairness notion).  In this paper, we mostly focus on share based fairness notions, and investigate the communication complexity of protocols that are acceptable under these notions.

\subsection{Fairness notions}
\label{sec:fairness}

A fairness notion is {\em feasible} if for every allocation instance, there is an allocation that is acceptable under the fairness notion. In this section we review the feasible fairness notions that will be considered in this paper. Some additional comments on these fairness notions can be found in Section~\ref{app:fair}.

\subsubsection{The MMS, and classes of valuations for which the MMS is feasible}
\label{sec:fairMMS}

The most commonly used share for settings with equal entitlements is the maximin share (MMS), introduced in~\cite{Budish11}. 

\begin{definition}
    \label{def:MMS}
    Let ${\cal{P}}_n$ denote the set of all partitions on $\items$ into $n$ sets. The maximin share (MMS) of an agent with valuation $v_i$ and entitlement $\frac{1}{n}$ is $\max_{P \in {\cal{P}}_n} \min_{S \in P} v_i(S)$, the value of the least valuable bundle in the best partition of $\items$ into $n$ bundles.  
\end{definition}

The MMS is not a feasible share for additive valuations~\cite{KPW18}. However, for classes of instances for which it is feasible, then in some rigorous sense, the MMS is the ``best possible" feasible share~\cite{BF22}. We list here classes of allocation instances for which MMS is feasible.

\begin{itemize}
    \item Setting with {\em identical valuations} (all agents have the same valuation function).
    \item Settings with {\em unit demand} valuations.
    \item Settings with {\em binary additive} (or {\em binary}, for short) valuations. Here, binary valuations refer to additive valuation functions in which the value of every single item is either~0 or~1. For binary valuations, Prop1 allocations are also Aprop, MXS and MMS.
    \item Settings with {\em 2-valued additive} (or {\em 2-valued}, for short) valuations. Here, 2-valued valuations refer to additive valuations in which there are two values, $a > b \ge 0$, such that the value of every single item is either~$a$ or~$b$. (In instances with several agents, the same $a > b \ge 0$ are used in all valuation functions. Feasibility of the MMS in this case is proved in~\cite{feige22}.) 
    \item Settings with two agents with additive valuations.
\end{itemize}

For $\rho \ge 0$, a $\rho$-MMS allocation is one in which each agent gets at least a $\rho$ fraction of her MMS. For additive valuations, the largest value of $\rho$ for which $\rho$-MMS is a feasible share is known to be at least $\frac{3}{4} + \frac{3}{3836}$~\cite{akrami2023breaking} and at most $\frac{39}{40}$~\cite{FST21}.

\subsubsection{Comparison based fairness notions: EF1, EFX and MXS}
\label{sec:fairComparison}

In this section we present comparison based fairness notions. We also present share based notions that are derived from the respective comparison based notions. 

\begin{definition}
    \label{def:EF}
    An allocation $A_1, \ldots, A_n$ is {\em envy-free} (EF) if for every two agents $i \not= j$ it holds that $v_i(A_i) \ge v_i(A_j)$. It is {\em envy-free up to one item} (EF1) if instead either $v_i(A_i) \ge v_i(A_j)$ or there is an item $e \in A_j$ for which $v_i(A_i) \ge v_i(A_j \setminus \{e\})$. It is {\em envy-free up to any item} (EFX) if instead either $v_i(A_i) \ge v_i(A_j)$ or for every item $e \in A_j$ it holds that $v_i(A_i) \ge v_i(A_j \setminus \{e\})$.
\end{definition}

(The notion EF1 is attributed to~\cite{Budish11}, and EFX to~\cite{CKMPSW19}.)

EF is not a feasible fairness notion. EF1 is feasible~\cite{LiptonMMS04}. It is an open question whether EFX is feasible, even just for additive valuations. See for example~\cite{AACGMM23}.

The {\em minimum EFX share} (MXS) is a share based fairness notions that is derived from EFX. It was introduced in~\cite{CaragiannisGRSV23}. 

\begin{definition}
 \label{def:MXS}
     The {\em minimum EFX share} (MXS) of an agent $i$ is the highest value $t$ that satisfies the following. In every allocation $A_1, \ldots, A_n$ in which bundle $A_i$ satisfies the EFX property  (for every $j \not= i$, either $v_i(A_i) \ge v_i(A_j)$, or $v_i(A_i) \ge v_i(A_j \setminus \{e\})$ for every item $e \in A_j$), it holds that $v_i(A_i) \ge t$. An MXS-allocation is one in which every agent gets at least her MXS.
 \end{definition}

An MXS allocation need not be an EFX allocation. {In fact, it need not even be an EF1 allocation. For example, for additive valuation functions with $m = 2n-1$ identical items, every allocation that gives each agent at least one item is MXS, but to be EF1, there is the additional constraint that no agent receives more than two items.} 

MXS is a feasible share. For additive valuations, this follows from an allocation algorithm of~\cite{BK20}, as observed in~\cite{CaragiannisGRSV23}. For general valuations, this was proved in~\cite{AR24}.

For additive valuations, it follows from~\cite{ABM18} that $\frac{4}{7}$-MMS $\le$ MXS $\le$ MMS.

\subsubsection{Additive valuations: Aprop, a relaxation of the proportional share}
\label{sec:fairAdditive} 

In this section we present fairness notions that are well suited for additive valuations. Consequently, we assume throughout this section that agents have additive valuations.

The {\em proportional share} of agent $i$ is $Prop(v_i, \items, \frac{1}{n}) = \frac{1}{n}v_i(\items)$. A {\em Prop-allocation} is an allocation that is acceptable under the fairness notion of Prop, namely, one in which each agent $i$ gets a bundle that she values at least as much as $Prop(v_i, \items, \frac{1}{n})$. 
As Prop is not a feasible share, we shall consider relaxations of Prop that are known to be feasible.

One type of relaxation is {\em proportional up to one item} (Prop1)~\cite{CFS17}. 

\begin{definition}
\label{def:Prop1}
 A Prop1-allocation $A_1, \ldots, A_n$ is one in which for each agent $i$, either $v_i(A_i) \ge \frac{1}{n} v_i(\items)$ (as in Prop-allocations), or there is an item $e \not\in A_i$ such that $v_i(A_i) + v_i(e) > Prop(v_i,\items, \frac{1}{n})$.    
\end{definition}

\begin{remark}
\label{rem:Prop1def}
    Our definition of Prop1 is slightly stronger than that of~\cite{CFS17}, as they only require  $v_i(A_i) + v_i(e) \ge Prop(v_i,\items, \frac{1}{n})$. In particular, if there are $n$ agents and $m=n$ identical items, a Prop1 allocation under the stronger definition requires that every agent gets an item, whereas under the weaker definition a Prop1 allocation may allocate all items to the same agent. Our positive results regarding Prop1 allocations trivially extend to the weaker definition, and our negative results can be adapted to hold also under the weaker definition. See Remark~\ref{rem:lowerProp1}.
\end{remark}

Prop1 is a feasible fairness notion: Prop1-allocations always exist. 

Another type of relaxation is to settle for (at least) a $\rho$-fraction of the $Prop(v_i, \items, \frac{1}{n})$ value, instead of the full $Prop(v_i, \items, \frac{1}{n})$. 
Unfortunately, no positive value for $\rho$ is feasible (e.g., if $m < n$). 
To address this concern, we consider the {\em truncated proportional share} (TPS), introduced in~\cite{BEF22BoBW}. 

\begin{definition}
    \label{def:TPS}
    Given a valuation $v_i$ and a number $n$ of agents, and item $e$ is {\em over-proportional} if $v_i(e) > \frac{1}{n}v_i(\items)$
    The value of the {\em truncated proportional share}, $TPS(v_i, \items, \frac{1}{n})$, is defined in a recursive manner. If there are no over-proportional items (no item $e$ with $v_i(e) > \frac{1}{n}v_i(\items)$), then $TPS(v_i, \items, \frac{1}{n}) = Prop(v_i, \items, \frac{1}{n})$. If there are over-proportional items, then let $e$ be an over proportional item (the end result will be the same whichever over-proportional item is picked), and then $TPS(v_i, \items, \frac{1}{n}) = TPS(v_i, \items \setminus \{e\}, \frac{1}{n-1})$. 
\end{definition}

We note that for additive valuations we have that $MMS \le TPS \le Prop$.

We shall consider $\rho$-TPS, with $\rho = \frac{n}{2n-1}$. This is the largest value of $\rho$ for which $\rho$-TPS is a feasible share~\cite{BEF22BoBW}.
No larger value of $\rho$ is possible when there are $m=2n-1$ identical items.

Both Prop1 and $\frac{n}{2n-1}$-TPS are feasible relaxations of Prop. Depending on the valuation functions, any one of them may be larger than the other.  For large $k$ and $n$, when there are $m = kn$ items each of value~1, the value of Prop1 is $k$, whereas $\frac{n}{2n-1}$-TPS is only roughly $\frac{k}{2}$. In contrast, when there is an over-proportional item, the value of Prop1 is~0, but $\frac{n}{2n-1}$-TPS is positive (it at least $n$ items have positive value).
{As} the two different ways in {which} Prop can be relaxed are incomparable, we require that both hold. 

\begin{definition}
    \label{def:Aprop}
    An allocation is {\em approximately proportional} (Aprop) if it is both Prop1 and $\rho$-TPS for $\rho = \frac{n}{2n-1}$.
\end{definition}

In Proposition~\ref{prop:MXSAprop} we show that every MXS allocation is also Aprop. In particular, this shows that Aprop is a feasible fairness notion.

\subsection{Communication complexity}
\label{sec:cc}

We study the communication complexity of achieving a fair allocation. Our setting is that of a publicly known allocation protocol, that is executed either collectively by the agents themselves, or by a third party. For every agent $i$, her valuation $v_i$ is known only to $i$. The protocol may send queries to $i$, so as to learn information about $i$ that is needed in order to produce a fair allocation. There are no restrictions regarding the type of queries that may be asked, and agents are assumed to answer queries truthfully. 
The {\em transcript} of the protocol is the list of replies sent by the agents. When the protocol ends, the rules of the protocol map the transcript into an allocation, and it is required that this allocation is fair. 
Our complexity measure is the length of the transcript, namely, the number of bits transmitted by the agents in reply to the queries. The queries themselves do not count as part of the communication complexity, as the protocol is publicly known, and hence the description of the queries is implicit from the description of the replies. 

A protocol may be randomized. In this case, the source of randomness that it uses is referred to as the {\em common random string} (CRS). The contents of the CRS is known to all agents, and can be thought of as selecting which deterministic protocol is being run (after fixing the contents of the CRS at random, the protocol is deterministic). For randomized protocol, we consider the expected communication complexity (expected length of the transcript), where expectation is taken over the choice of the CRS. 

For a class $C$ of valuations ({\em UD} denotes Unit Demand, {\em Bin} denotes Binary, {\em 2V} denotes 2-Valued, {\em Add} denotes additive) and a fairness notion $F$ (such as MMS, Aprop, MXS) we express communication complexity as a function of $m$ (number of items), and $n$ (number of agents). For a choice of $m$ and $n$, the upper bounds given apply to all input instances with with $m$ items and $n$ agents and valuations from the class $C$, whereas lower bounds mean that for every protocol, at least one allocation instance requires communication complexity at least as high as the lower bound.  

We shall use the notation $CC(F,C)$ to denote communication complexity for deterministic protocols, and $RCC(F,C)$ to denote expected communication complexity for randomized protocols. Dividing the communication complexity by $n$, we get the {\em average} communication complexity per agent. This will be denoted by $ACC(F,C)$ and $ARCC(F,C)$ (which equal $\frac{1}{n}\cdot CC(F,C)$ and $\frac{1}{n}\cdot RCC(F,C)$, respectively). 

\subsubsection{Description complexity}
\label{sec:dc}

In addition to communication complexity (CC), it will be instructive to consider what we refer to as the {\em description complexity} (DC), which measures how many bits are required in order to describe the fair allocation. In the language of protocols, (randomized) description complexity is the (expected) communication complexity of a (randomized) fair allocation protocol in which there is a special agent that knows the valuation functions of all agents, and all queries are directed to the special agent. As queries can be arbitrary, a query may simply ask the special agent for the final fair allocation (the agent has sufficient information in order to compute it), and so the (expected) length of the reply of the special agent serves as the length of the description of the allocation. 

Description complexity serves as a lower bound on communication complexity. Description complexity is tightly related to the notion of {\em hitting sets}.

\begin{definition}
    \label{def:hittingSet}
    Given a fairness notion $F$, a class $C$ of valuations, $n$ and $m$, a set $S$ of allocations is referred to as a {\em hitting set} if for every allocation instance with $m$ items and $n$ agents with valuations from the class $C$, at least one of the allocations in $S$ is acceptable under $F$. 
\end{definition}

It is not hard to see that the description complexity is simply the logarithm of the smallest hitting set. As the set of all allocations is a hitting set and has size $n^m$, it follows that description complexity is never larger than $m \log n$.

Randomized description complexity is tightly related to the notion of fractional hitting sets in which each member of $S$ is given a non-negative weight, and the weights of those members of $S$ that are acceptable for any given instance must sum up to at least~1. It can be shown that up to constant factors, the randomized description complexity is the logarithm of the weight of the minimum weight fractional hitting set.


For description complexity, we use the notation $DC$, $RDC$, $ADC$ and $ARDC$, for deterministic, randomized, average, and average randomized description complexity, respectively.

{\bf Identical valuations.} In allocation instances in which all agents have the same valuation function, there is no distinction between communication complexity and description complexity, because every agent knows the valuations of all agents. 

\subsection{Some intuition}
\label{sec:intuition}

In this section we make a few observations that may help the reader to gain intuition as to what kind of results one can expect to have in our setting. 

{\bf Conventions, applicable for all sections of this paper.} For simplicity, we omit floor and ceiling notation from expressions such as $\lceil \log m \rceil$. Also, we assume unless explicitly stated otherwise that $m \ge 2n$, so expressions such as $\log \frac{m}{n}$ have value of at least~1.

Recall that we are dealing with item based valuations. Hence, if each item can have one of $P$ possible known values (e.g., an integer value in the range $[0, P-1]$), then an agent may completely describe her valuation function by communicating $m \log P$ bits. Given the full description of all valuation functions, the protocol can compute a fair allocation. (We are not concerned here with the computational complexity of computing a fair allocation, only with the communication complexity.) It follows that for classes $C$ of item based valuations and any feasible fairness notion $F$, it holds that $ACC(F,C) \le m \log P$. In particular, for 2-valued valuations we have that $ACC(F,2V) \le m$. 

For the fairness notions that we consider, we may expect lower communication complexity. For example, it is well known that for additive valuations, the {\em round robin} protocol (in which agents are visited in a round robin fashion, and each agent in her turn selects the item most valuable to her among those that still remain) produces an EF1 allocation. As an item can be specified using $\log m$ bits, we get that $ACC(EF1,Add) \le \frac{m}{n}\log m$. This improves over the bound of $ACC(EF1,Add) \le m \log P$ when $\log m < n \log P$. Moving to unit demand valuations, it suffices to do just one round of the round robin algorithm (giving all remaining items to one agent arbitrarily) in order to get allocations that are simultaneously MMS and EF1, establishing that $ACC(MMS,UD) \le \log m$.

Description complexity will help us gauge how much the fact that the input (the valuation functions) is distributed among several agents contributes towards the communication complexity. As an allocation can be complexity specified using $m \log n$ bits (stating for each item which agent receives it), it follows that $DC(F,C) \le m \log n$, for every feasible fairness notion $F$ and every class $C$ of valuations. 

Often, the description complexity is even smaller. In particular, this happens when $n$ is much smaller than $m$, for fairness notions that always have {\em contiguous} acceptable allocations (arranging the items on a line from $e_1$ to $e_m$, each agent gets an interval of consecutive items). A contiguous allocation can be specified using only $\log (\binom{m}{n-1} \cdot n!) = O(n \log m)$ bits. (The $\binom{m}{n-1}$ term comes from selecting the endpoints of the first $n-1$ intervals, and the term $n! \simeq (\frac{n}{e})^n$ comes from specifying which agent gets which interval.) Note that $n\log m$ may be much smaller than $m \log n$, as it may be that $n$ is constant and $m$ is arbitrarily large (we always assume that $m > n$). 

For additive valuations, Prop1 is known to have contiguous allocations~\cite{Suksompong19}. Hence, we have that $DC(Prop1,Add) \le O(n \log m)$, and $ADC(Prop1,Add) \le O(\log m)$. For every class $C$ of valuations, it is known that $EF1$ has contiguous allocations~\cite{BiloCFIMPVZ22, Igarashi23}. Hence we also have that $DC(EF1,C) \le O(n \log m)$, and $ADC(EF1,C) \le O(\log m)$.

Given the above background, here are some questions to keep in mind.

\begin{enumerate}
    \item Beyond unit demand, which classes of valuations have average (randomized) communication complexity of $O(\log m)$, for natural fairness notions? Can we go even below $\log m$?
    \item Which fairness notions and classes of item-based valuations have average (randomized) communication complexity of $\Omega(m)$? 
    \item For which fairness notions and classes of valuations is the communication complexity similar in magnitude to the description complexity? Are there examples of large gaps between the two? 
    \item Does randomization help? Are there examples in which randomized communication complexity is smaller than the deterministic communication complexity? Could it be that the randomized communication complexity is even smaller than the (deterministic) description complexity?
    \item Identify locations of ``jumps" in the complexity: a relatively small extension of a class of valuations that leads to a big increase in the communication complexity (for some fairness notion), or a relatively small strengthening of a fairness notion that leads to a big increase in the communication complexity (for some class of valuations).
\end{enumerate}

\subsection{Our main results}
\label{sec:results}

In this section we state our main results. They are arranged in several subsections.

\subsubsection{MMS allocations}
\label{sec:ResultsMMS}

For several classes of valuation functions for which the MMS is feasible, we achieve tight or nearly tight characterization of the randomized communication complexity.

\begin{theorem}
    \label{thm:MMS}
    The average randomized communication complexity of MMS allocations is $\Theta(1)$ for unit demand valuations, $\Theta(\log \frac{m}{n})$ for binary additive valuations, and $\Omega(\frac{m}{n})$ for 2-valued additive valuations. For more details, see Table~\ref{tab:MMS}.
\end{theorem}

\begin{table}[hbt!]
        \centering
\begin{tabular}{ | m{2cm} || m{2.8cm}| m{2.8cm} | m{2.8cm} |} 
 \hline
 \multicolumn{4}{|c|}{randomized communication complexity for MMS allocations} \\
 \hline
 Valuation \newline class & RCC & RCC for \newline identical \newline valuations & average \newline randomized \newline complexity \\ 
 \hline
  \hline
 Unit \newline Demand & $\Theta(n)$ & $\Theta(n)$  & $\Theta(1)$ \\
 \hline
 Binary & $\Theta(n \log \frac{m}{n})$ & $\Theta(n \log \frac{m}{n})$   & $\Theta(\log \frac{m}{n})$  \\ 
 \hline
 2-valued  & $O(m \log n)$ & $\Theta(m)$ & $\Omega(\frac{m}{n})$ \\ 
 \hline
\end{tabular}
        \caption{Dependence of average expected communication complexity on class of allocation instances. Expectation is taken over randomness of the protocol. Average is taken over the agents (dividing the communication complexity by $n$). For 2-valued valuations, the upper bound holds even or deterministic protocols, showing that $CC(MMS,2V) \le O(m\log n)$.}
        \label{tab:MMS}
\end{table}

We provide a few comments on Theorem~\ref{thm:MMS}. For Unit Demand valuations and Binary valuations, we show the existence of randomized protocols with surprisingly low expected communication complexity. On average, each agent sends only a constant number of bits. For Unit Demand valuations this holds regardless of the values of $m$ and $n$, and for Binary valuations, this holds when $m = O(n)$. In contrast, for 2-valued valuations, if the number of agents is constant, then on average each agent needs to send $\Omega(m)$ bits, which up to a constant factor suffices in order to describe her full valuation function (being 2-valued, it can be fully described by $m$ bits). 

For 2-valued valuations, the precise values of $a > b \ge 0$ makes a huge difference to the average randomized communication complexity. Fixing $a = 1$, if $b = 0$ we are in the binary case, where $O(\log \frac{m}{n})$ suffices, whereas for all $0 < b \le \frac{1}{m}$ there is an exponential jump, to $\Omega(\frac{m}{n})$, even for identical valuations.

For the classes of valuations that we considered, there is not much of a different between randomized communication complexity and randomized description complexity. The (randomized) description complexity cannot be lower than (randomized) communication complexity for identical valuations, and our lower bounds for identical valuations match up to small multiplicative factors (constant, or $O(\log n)$) our upper bounds for the non-identical setting.


For deterministic communication complexity, our bounds have more slackness than for randomized communication complexity.

\begin{theorem}
    \label{thm:MMSdet}
    The deterministic communication complexity of MMS allocations is $O(n ( \log n + \log\log \frac{m}{n}]))$ for unit demand valuations, $O(n \log m \log n)$ for binary additive valuations, and $O(m \log n)$ for 2-valued additive valuations. For more details, see Table~\ref{tab:MMSdet}.
\end{theorem}

\begin{table}[hbt!]
        \centering
\begin{tabular}{ | m{2cm} || m{4cm}  | m{4cm} |} 
 \hline
 \multicolumn{3}{|c|}{deterministic communication complexity for MMS allocations} \\
 \hline
 Valuation \newline class & CC & CC for \newline identical \newline valuations \\ 
 \hline
  \hline
 Unit \newline Demand & $O(n (\log n + \log\log \frac{m}{n}))$ & $\Theta(n + \log\log m)$   \\
 \hline
 Binary & $O(n \log m \log n)$ & $\Omega(n \log \frac{m}{n})$      \\ 
 \hline
 2-valued  & $O(m \log n)$ & $\Theta(m)$  \\ 
 \hline
\end{tabular}
        \caption{Dependence of deterministic communication complexity on class of allocation instances.}
        \label{tab:MMSdet}
\end{table}

We provide a few comments on Theorem~\ref{thm:MMSdet}. For Unit Demand valuations and for Binary valuations, our upper bounds pay some extra log factors compared to the randomized communication complexity. For identical unit demand valuations, our lower bound of $\Theta(n + \log\log m)$ (see Table~\ref{tab:MMSdet}) is stronger than the lower bound in the randomized case. Note that for constant $n$, it implies a lower bound of $\log \log m$ on the description complexity of MMS allocation for Unit Demand valuations, whereas the randomized communication complexity of finding such allocations is $O(1)$ (by Theorem~\ref{thm:MMS}). This illustrates that randomized communication complexity can be lower than deterministic description complexity (but of course, it cannot be lower than randomized description complexity).

MMS allocations are not feasible for general additive valuations. Some sub-classes of additive valuations (binary, 2-valued) for which MMS is feasible are addressed above. The next theorem addresses additional special cases in which MMS is feasible. 

\begin{theorem}
    \label{thm:MMSadditive}
    The communication complexity of MMS allocations for additive valuations is $\Theta(m \log n)$ in the special case of identical valuations, and $\Theta(m)$ in the special case of two agents. See Table~\ref{tab:MMSadditive}.
\end{theorem}

\begin{table}[hbt!]
        \centering
\begin{tabular}{| c || c | c |}
 \hline
 \multicolumn{3}{|c|}{MMS allocations for additive valuations} \\
 \hline
 Special 
 case & Deterministic & Randomized \\ 
 \hline
  \hline
 Identical 
 valuations & $\Theta(m \log n)$ & $\Theta(m \log n)$  \\ 
 \hline
 Two 
 agents & $\Theta(m)$ & $\Theta(m)$   \\
 \hline
\end{tabular}
        \caption{Communication complexity of MMS allocations for additive valuations, in two special cases in which MMS allocations are known to be feasible.}
        \label{tab:MMSadditive}
\end{table}

We provide a few comments on Theorem~\ref{thm:MMSadditive}. The upper bounds implied by this theorem are straighforward. For 2-agents, an $O(m)$ upper bound is implied by the well known {\em cut and choose} protocol, whereas for identical valuations the upper bound is respected when one of the agents simply outputs her own MMS partition.  For the two agent case, the $\Omega(m)$ lower bound is simply that of the 2-valued case in Theorem~\ref{thm:MMS}. Consequently, the main content of Theorem~\ref{thm:MMSadditive} is in the lower bound for identical valuations. Equivalently, this is a lower bound on the description complexity of the MMS partition of a single agent.

\subsubsection{General additive valuations}
\label{sec:ResultsAdditive}

For additive valuations in general,  MMS allocations need not exist~\cite{KPW18}. Here we consider other fairness notions that are feasible for additive valuations.

Theorem~\ref{thm:MMS} shows (among other things) a lower bound of $\Omega(m)$ on the randomized communication complexity of MMS allocations for 2-valued valuations, which are a subclass of additive valuations. The lower bound holds also for identical valuations. The exact same example that proves the lower bound applies also to a weaker share-based fairness notion, MXS, that is feasible for additive valuations.

\begin{corollary}
    \label{cor:binaryMXS}
    For any number $n \ge 2$ of agents, the randomized communication complexity for finding MXS allocations is at least $\Omega(m)$, even for identical 2-valued valuations (and hence, also more generally for additive valuations). 
\end{corollary}

\begin{remark}
    \label{rem:EFX}
    Corollary~\ref{cor:binaryMXS} holds also for other fairness notions, including EFX (as every EFX allocation is MXS, note that EFX allocations exist for 2-valued valuations) and EQX (see definition~\ref{def:EQX}). See Section~\ref{app:EQX} for more details. For EFX, the proof extends also to binary valuations. 
\end{remark}

Corollary~\ref{cor:binaryMXS} shows that for a constant number of agents, computing an MXS allocation requires average randomized communication complexity of $O(m)$ per agent. The next theorem shows that for a weaker share-based fairness notion, Aprop (which combines Prop1 and $\frac{n}{2n-1}$-TPS), the average randomized communication complexity improves exponentially, to $O(\log m)$.

\begin{theorem}
    \label{thm:Aprop}
    For additive valuations, the deterministic communication complexity of Aprop allocations is $O(n \log m \log n)$, and the randomized one is $O(n \log m)$. The randomized communication complexity is at least $\Omega(n \log \frac{m}{n})$, even for just Prop1 allocations and identical binary valuations. See Table~\ref{tab:Aprop}.
\end{theorem}

\begin{table}[hbt!]
        \centering
\begin{tabular}{| c || c | c |}
 \hline
 \multicolumn{3}{|c|}{Aprop allocations for additive valuations} \\
 \hline
 & Deterministic & Randomized \\ 
 \hline
  \hline
 In general & $O(n\log m \log n)$ & $O(n \log m)$   \\
 \hline
 Identical 
 valuations & $\Theta(n \log \frac{m}{n})$ & $\Theta(n \log \frac{m}{n})$  \\ 
 \hline
\end{tabular}
        \caption{Communication complexity of Aprop allocations for additive valuations. Recall that description complexity cannot be smaller than communication complexity for identical valuations.}
        \label{tab:Aprop}
\end{table}

The lower bounds in  Theorem~\ref{thm:Aprop} for identical valuations are an immediate corollary of Theorem~\ref{thm:MMS}, because for binary valuations there is no distinction between (our definition of) Prop1 allocations and MMS allocations (see Remark~\ref{rem:Prop1def}). Consequently, the main content of Theorem~\ref{thm:Aprop} is in the upper bounds.

\subsubsection{EF1 allocations}
\label{sec:EF1results}

For additive valuations,
it is well known that the Round Robin protocol produces EF1 allocations. It is also known that there always is a contiguous EF1 allocation~\cite{BiloCFIMPVZ22, Igarashi23}. These results easily imply the following proposition.

\begin{proposition}
    \label{pro:EF1}
    For additive valuations, the communication complexity of EF1 allocations is $O(m \log m)$ and the description complexity is $O(n \log m)$. For identical valuations, the communication complexity is $O(n \log \frac{m}{n})$.
\end{proposition}

Our main result for EF1 allocations concerns unit demand valuations.

\begin{theorem}
    \label{thm:unitDemandEF1}
    For unit demand valuations, the randomized communication complexity of EF1 allocations is $O(n \log n)$, and the randomized description complexity is $\Omega(n \log n)$. Hence, the average randomized communication complexity per agent is $\Theta(\log n)$.
\end{theorem}

The lower bound in Theorem~\ref{thm:unitDemandEF1} is for instances in which different agents have different valuations functions. As the lower bound holds even when $m = O(n)$, having different valuations is necessary, because for identical valuations Proposition~\ref{pro:EF1} gives an upper bound of $O(n)$ is this case. 

\subsubsection{Computational complexity}
\label{sec:computational}

All our lower bounds hold even if agents are allowed unbounded computation time.
Almost all our upper bounds stated above are via allocation algorithms that can be implemented in polynomial time. There are two exceptions. One is in Theorem~\ref{thm:MMSadditive}, and is unavoidable, because even for identical additive valuations, computing MMS allocations is NP-hard. The other concerns Proposition~\ref{pro:EF1}, and specifically the description complexity of $O(n \log m)$ for EF1 allocations for additive valuations. It is based on the existence of contiguous EF1 allocations, but the known proof that such allocations exists~\cite{Igarashi23} does not give a polynomial time algorithm (it is based on Sperner's lemma).


\subsection{Related work}
\label{sec:related}



For general background on communication complexity, see for example~\cite{KN97}. In the context of allocation problems, communication complexity questions are studied in~\cite{NS06} (and also in more recent work, such as~\cite{EFNTW19}), though there the setting is different than ours. These works prove exponential lower bounds on the communication complexity of achieving an allocation that maximizes (or approximately maximizes) social welfare, and crucially for these lower bounds, valuations of agents are not additive.  

{For a recent survey on fair division of indivisible goods, see~\cite{amanatidis2023fair}. Quoting from that survey:  {\em it is an important research direction to explore how much information about the valuations of the agents is sufficient to design algorithms with strong fairness guarantees.}}

The communication complexity of fairly allocating indivisible goods was studied in~\cite{PR20a}. Fair allocation problems were classified  based on five attributes: number of agents $n$; class of valuation functions:  submodular, subadditive, or general; fairness notion: EF or Prop; the allowed approximation ratio; deterministic versus randomized communication complexity. For each setting of the attributes, the question asked is what is the communication complexity of determining whether an acceptable allocation exists -- is the communication complexity polynomial or exponential? The case of additive valuation was not considered in that work, because in that case communication that is polynomial in $m$, $n$ and $\max_{\{i,e\}}[\log v_i(e)]$ suffices in order to completely learn the valuation functions of all agents.

The communication complexity of fairly allocating a divisible good $G$ was studied in~\cite{BN19}. To allow for short answers, it is assumed that for every agent $i$ it holds that $v_i(G) = 1$, and that an allocation is acceptable to agent $i$ if she receives a value of at least $\frac{1}{n} - \epsilon$ (one should think of $\epsilon$ as being smaller than $\frac{1}{n}$). The lower bounds shown in that work are of the form $O(\log(\frac{1}{\epsilon}))$, but without explicit reference to the dependence on $n$. One can translate these lower bounds from the divisible case to the indivisible one, by taking $\epsilon < \frac{1}{3m}$. This will imply a lower bound of $\Omega(\log m)$ on the communication complexity of Aprop allocations. 
Our lower bound in Theorem~\ref{thm:Aprop} is as good when $n$ is constant, and much better as $n$ grows. 
Upper bounds in the divisible case need not imply upper bounds in indivisible case, as they might correspond to splitting items.  

There is a class of works related to communication complexity problems, and it refers to query complexity. In that setting, there is a central authority that attempts to compute a fair allocation, but does not know the valuations of the participating agents. It can attempt to gain information about these valuations through queries to the agents. Typically, these queries are limited to be of a special type. A common example is {\em value queries}, in which the central authority specifies a bundle of items, and the agent replies with its value. The query complexity of a protocol is the number of queries that are asked, whereas the sizes of queries and answers are typically not a consideration. Hence query complexity might be smaller than the communication complexity (if queries have long answers), and might be larger than the communication complexity (due to the fact that in the query model, all communication must be by answers to queries of a given type). 

The query complexity of fairly allocating indivisible goods was studied in several works. In~\cite{PR20b}, among additional results, it is shown that if agents have submodular valuations, then finding an EFX allocation requires an exponential number of value queries, even if there are only two agents. In~\cite{OPS21} it is shown that for additive valuations and two agents, $\Omega(m)$ value queries are required in order to compute EFX allocations. Our Corollary~\ref{cor:binaryMXS} shows an $\Omega(m)$ communication complexity lower bound for MXS allocation (and hence also for EFX allocations), and moreover, our lower bound holds also for randomized protocols. Another negative result of~\cite{OPS21} is a lower bound of $\Omega(n\log m)$ on the number of value queries required by a deterministic algorithm to compute an EF1 allocation, when valuations are binary and $m > n^{1 + \Omega(1)}$.  Our Theorem~\ref{thm:unitDemandEF1} shows an $\Omega(n \log n)$ communication complexity lower bound for EF1 allocations with binary valuations even when $m = O(n)$, and moreover, our lower bound holds also for randomized protocols and for unit demand valuations. On the positive side, it is shown in~\cite{OPS21} that EF1 allocations (and hence also Prop1 allocations) can be computed with $O(\log m)$ queries for additive valuations and $n \le 3$, and $O((\log m)^2)$ queries for general monotone valuations and three agents. For any number of agents, the results of~\cite{LiptonMMS04} imply that an EF1 allocation can be computed in a number of queries that is linear in $m$ and polynomial and $n$. {The same applies to MXS allocations for additive valuations, by the algorithm of~\cite{BK20}.}

The work in~\cite{bu2024fair} considers a query model that allows only comparison queries. A comparison query to an agent $i$ is a pair of bundles $Q_1$ and $Q_2$, and the reply is~1 if $v_i(Q_1) \ge v_i(Q_2)$, and~0 otherwise. It is shown in~\cite{bu2024fair} that $O(n^4 \log m)$ comparison queries suffice in order to find allocations that are simultaneously Prop1 and $\frac{1}{2}$-MMS. As replies to comparison queries are only single bits, this implies that the communication complexity of finding such allocations is no worse than $O(n^4 \log m)$. Our Theorem~\ref{thm:Aprop} gives stronger upper bounds, showing that $CC(Aprop) \le O(n\log n \log m)$, but our protocol uses queries that are not comparison queries.
Our Theorem~\ref{thm:Aprop} gives an $\Omega(n \log \frac{m}{n})$ lower bound for Prop1 allocations. This implies a similar lower bound on comparison queries, which is stronger than the $\Omega(\log m)$ lower bound shown in~\cite{bu2024fair}. (Note however that the focus in~\cite{bu2024fair} is on instances with a constant number of agents. In that special case, the only distinction between these lower bounds is that our holds also for randomized protocols.)

The query complexity of fairly allocating a divisible good has received significant attention.  
A common query model is the Robertson-Webb model (RW) with two types of queries (value of interval, location of cut). Our upper bounds in Theorem~\ref{thm:Aprop} make use of algorithms of~\cite{EP84} designed for this model. {In contrast,} the known lower bound of $\Omega(n\log n)$ queries for deterministic protocols that produce proportional allocations in this model~\cite{EP11} {is} not useful for us, for two reasons. One is that {it uses} in an essential way the query model. In particular, the proof uses the assumption that to output a minimal (in the sense that the algorithm does not have sufficient information in order to trim it) collection of $k$ intervals of a certain desired density (value over length), $k$ queries are required. Hence, one can afford $k$ additional queries to learn the value of each of these intervals, and pick the best one. However, in general communication complexity models, one can in a single query ask about the density of a union of any number of intervals. The other reason why the lower bound {is} not useful for us is because the class of valuations, when translated to our setting of divisible items, requires $m > n^{1 + \Omega(1)}$, and for this range of parameters our lower bound of $n\log \frac{m}{n}$ is at least as strong, and holds even for randomized protocols. In the other direction, our lower bound does not imply a query lower bound in the divisible case, because we count bits in the answers, whereas the query lower bound allows for answers with arbitrary precision. The randomized query complexity of achieving an allocation that is approximately proportional (in some sense) was shown to be only $O(n)$~\cite{EP06}. However, it does not seem to us that the techniques used in that work are useful in our context (of indivisible items, and the kind of approximations of proportionality that we aim to achieve). The query complexity for allocation of a divisible good in which fairness notions need to be satisfied up to an accuracy parameter $\epsilon$ is studied in~\cite{BN22}.









\subsection{Preliminaries and assumptions that we make}
\label{sec:preliminaries}

We always assume that the number of items is larger than the number of agents, $m > n$. This is because if $m \le n$, every allocation that gives each agent at most one item satisfies all fairness properties considered in this paper. We shall further assume that $m \ge 2n$ (the constant~2 can be replaced by any other constant greater than~1). This serves two purposes. One is to ensure than $\log \frac{m}{n} \ge 1$ (avoiding the need to write $\max[1, \log \frac{m}{n}]$ in our bounds). The other is because in some cases, the case of $m$ very close to $n$ (e.g., $m = n+1$) really involves different bounds. See for example Proposition~\ref{pro:UDgap}.

For additive valuations, we assume that there are no over-proportional items, namely, for every agent $i$ and item $e$, $v_i(e) \le \frac{1}{n}v_i(\items)$. This assumption can be made without loss of generality, because for every over-proportional item, the agent may pretend that its value is in fact smaller, equal to the TPS of the agent. Every allocation that satisfies any of our fairness notions (Prop1, Aprop, MXS, MMS) with respect to this modified valuation satisfies them with respect to the original valuation. Consequently, if valuations are scaled so that $v_i(\items) = n$ (all our fairness notion are unaffected by multiplicative scaling of valuations), then no item has value above $1$.





A simplifying convention that we make is to ignore rounding effects. That is, if an agent has $k$ publicly known alternatives to choose from, we say {that} transmitting the chosen alternative takes $\log k$ bits, without rounding up to the nearest integer. This simplifying convention affects the communication complexity by at most a constant multiplicative factor, and this constant is not far from~1. 

When encoding integer values, we use an encoding scheme in which each integer $k \ge 0$ is encoded by $O(\log k)$ bits. (For $k \in \{0,1\}$ we interpret the notation $O(\log k)$ as meaning $O(1)$.) For example, this can be done by encoding the integer in binary, followed by a special symbol to denote ``end of message".  Then this ternary alphabet can be replaced by a binary one, encoding every ternary character by two bits.

The term {\em $F$ bundle} for agent $i$ for a fairness notion $F$ (Prop1, Aprop, MXS) refers to a set of items that is acceptable to $i$ under $F$. 


\section{Our approach for proving lower bounds}
\label{sec:lower}




Almost all our lower bounds will be proved for randomized description complexity RDC, and hence apply also to CC, RCC and DC.

In our communication complexity lower bounds, we shall make extensive use of the following lemma, a variation of the well known minimax principle of Yao~\cite{Yao77}.

\begin{lemma}
\label{lem:lowerBounds}
    For given $n$ and $m$, let $C_{n,m}$ be a class of allocation instances with $n$ agents and $m$ items, and let $F$ be a desired fairness property. Let $0 < p < 1$ be such that for every allocation $A$, the probability that $A$ is acceptable under $F$  is at most $p$. Here, the probability is taken over the choice of input instance chosen uniformly at random from  $C_{n,m}$. Then, the randomized description complexity for allocations acceptable under $F$ for allocation instances in $C_{n,m}$ is $\Omega(\log \frac{1}{p})$. That is, $RDC(F) \ge \Omega(\log \frac{1}{p})$. \end{lemma}

\begin{proof}
Consider a randomized algorithm $P$ that given the valuations of all agents, outputs an allocation acceptable under $F$. Let $T$ denote the expected number of bits output $P$. Here, expectation is taken over the contents of the random string (RS) used by $P$, whereas the input instance is a worst case input instance from $C_{n,m}$. 
By Markov's inequality, with probability at least half over the choice of RS, the output of the algorithm is at most $2T$ bits. {This last statement holds for every input instance from $C_{n,m}$.} Hence, there is a particular setting $r$ of the RS under which, for at least half the possible inputs from $C_{n,m}$, the output is of length $2T$ or less. As at most $2^{2T+1}$ different allocations can be encoded with at most $2T$ bits, there is some output $\tau$ such that a fraction of at least $2^{-2T-2}$ of the input instances from $C_{n,m}$ produce the output $\tau$ when the contents of the RS is $r$. As the combination of $r$ and $\tau$ uniquely determine the output allocation, there is an allocation that is acceptable according to $F$ for a fraction of at least $2^{-2T-2}$ of the input instances in  $C_{n,m}$. It follows that $p \ge 2^{-2T-2}$, implying that $T = \Omega(\log \frac{1}{p})$, as desired.    
\end{proof}

\section{Unit demand valuations}

In this section we consider Unit Demand valuations, proving the parts that refer to unit demand in Theorems~\ref{thm:MMS} and~\ref{thm:MMSdet},  and proving Theorem~\ref{thm:unitDemandEF1} and Proposition~\ref{pro:UDgap} (stated in Section~\ref{sec:openCCDC}).  

For the purpose of MMS allocations, we can (and will) replace unit demand valuations by binary valuations, in which for each agent exactly $n$ items have value~1 (these correspond to her $n$ top items, breaking ties arbitrarily), and the remaining items have value~0. An MMS allocation with respect to these binary valuations needs to give each agent at least one item of value~1, and hence at least one of the top $n$ items in the corresponding unit demand valuation, implying an MMS allocation for the unit demand case.

\subsection{Warm-up -- two unit-demand agents}
\label{sec:unitDemand}

We consider a setting with two agents, $a_1$ and $a_2$, and a set $\items$ of $m$ items $\{e_1, \ldots, e_m \}$. Each agent $a_i$ has a binary valuation $v_i$ in which exactly two items, referred to as the distinguished items,  have value~1. Hence an MMS allocation is one in which each agent receives a bundle that contains at least one of her distinguished items. 



The most naive allocation protocol is for each agent to submit her valuation function. Then, with full knowledge of these valuation functions, it is easy to output an Aprop allocation. For example, $a_1$ receives her first distinguished item, and $a_2$ receives all other items. As specifying a valuation function can be done using $2\log m$ bits (it suffices to specify the distinguished items), the communication complexity of the naive protocol is $4\log m$.

Clearly, the naive protocol is not optimal. A simpler and more efficient protocol is for $a_1$ to announce one of her distinguished items, and then to output the allocation that gives $a_1$ the announced item, and give $a_2$ the remaining items. This uses only $\log m$ bits of communication. 

The above protocol is optimal (perhaps up to an $O(1)$ additive term) among all protocols in which an observer learns the identity of at least one of the distinguished items (details omitted). However, can protocols that do not reveal such information be more efficient?  

Here is a more efficient protocol. Write the names of items in binary. Then the two distinguished items for $a_1$ differ in at least one bit location. Agent $a_1$ can announce one such bit location $\ell$. This partitions $\items$ into two parts $M_0$ and $M_1$, where $M_j$ contains those items whose name have $j$ in bit location $\ell$. Now $a_2$ selects one of these parts for herself (at least one of them contains a distinguished item for $a_2$), and $a_1$ gets the remaining part. The communication complexity is only $\log\log m + O(1)$ (the message of $a_1$ takes $\lceil \log\log m \rceil$ bits, whereas the message of $a_2$ takes only one bit).

The above protocol is optimal (perhaps up to an $O(1)$ additive term), even in terms of description complexity. Consider an arbitrary {\em hitting set} $H$ for MMS allocations (a set of allocations such that for every input instance from our class, at least one allocation in $H$ is MMS), and let $A_1, \ldots, A_{|H|}$ denote the allocations that it contains. Use this hitting set as an encoding scheme for names of items as follows. Each item $e_j$ receives a new $|H|$-bit name, where bit $\ell$ is~0 or~1 depending on whether $e_j$ is given to $a_1$ or to $a_2$ in allocation $A_{\ell}$. If $H$ is a hitting set, the names of all items are distinct, because if two items have the same name, these two items might be the distinguished items for both agents, and then no allocation in $H$ is MMS for this choice of valuation functions. If follows that $|H| \ge \log m$, and that the description complexity is at least $\log\log m$. 

Let us turn now to randomized communication complexity. It will convenient for us to assume here that $m$ is a power of~2, but we remark that the results can be extended to all values of $m$. 
Consider a random permutation over item names. 
Now the expected lowest bit location $\ell$ on which the names of $a_1$'s two distinguished items differ is~2. Moreover, encoding $\ell$ in unary (as $0^{\ell-1}1$) takes $\ell$ bits. Hence in expectation, $a_1$ transmits only two bits, and the randomized communication complexity is only~3 bits.

It may sound surprising that~3 bits of information suffice on average in order to completely specify a fair allocation. This is perhaps less surprising once one observes that in the setting considered in this section, a uniformly random allocation has probability at least $\frac{1}{2}$ of being fair. As there are so many fair allocations, finding one of them should not be difficult. 



Using the above discussion, we have established the following theorem.

\begin{theorem}
\label{thm:unitDemand}
For instances with two agents and unit demand valuations, the communication complexity for finding MMS allocations 
is as follows.
\begin{itemize}
    \item The deterministic communication complexity is $\log\log m + O(1)$.
    \item The randomized communication complexity is at most~3.
    \item The description complexity is a least $\log\log m$.
\end{itemize}

Moreover, in both the deterministic and the randomized protocols establishing the upper bounds on deterministic and randomized communication complexity, it is the case that agent $a_2$ gets her top valued item (and not just her MMS).

\end{theorem}

\begin{remark}
\label{rem:randomizedQueryComplexity}
It may be instructive to compare bounds for communication complexity and value-query complexity for instances with two agents and binary valuations with two items of value~1. Theorem~\ref{thm:unitDemand} shows that the deterministic communication complexity is $\Theta(\log\log n)$. In contrast, the results of~\cite{OPS21} imply that the deterministic value-query complexity is $\Theta(\log n)$. 
Theorem~\ref{thm:unitDemand} shows that the randomized communication complexity is $O(1)$. The same holds also for randomized value-query complexity, via similar arguments. This illustrates that randomized query complexity can be significantly smaller than deterministic query complexity.
\end{remark}

\subsection{Identical unit demand valuations}

Here we consider the case that there are $n$ agents with identical unit demand valuations, 
or equivalently (for our fairness notions), in which all agents have the same binary valuation $v$, with $v(\items) = n$. Moreover, we shall assume that $m \ge 2n$, so as not to bother with edge cases such as $m = n+1$.

Recall that for identical valuations, there is no distinction between description complexity and communication complexity. 

As we shall later see, the proof of {Theorem~\ref{thm:binaryLower}} implies a lower bound of $\Omega(n)$ for the randomized description complexity of such instances. The proof of Theorem~\ref{thm:unitDemand} implies a lower bound of $\log\log m$ on deterministic description complexity, even when there are only two agents. This easily extends to $\log\log (m - n + 2) = \Omega(\log\log m)$ (when $m \ge 2n$) for any number of agents, by revealing the location of $n-2$ of the items that have value~1. Hence we have a lower bound of $\Omega(n + \log\log m)$ for deterministic description complexity.

Interestingly, there is also a matching upper bound. This is because there is a family of $2^{O(n)}(\log m)^{O(1)}$ allocations (the logarithm of this is $O(n + \log\log m)$), such that for every instance of the above form, at least one allocation in the family is an MMS allocation. Such a family is given by a natural correspondence with $n$-perfect families of Hash functions form $[m]$ to $[n]$. For more details on these families of hash functions and their construction, see~\cite{SchmidtS90, AlonYZ95}.

Randomized protocols can shave off the $\log\log m$ factor. Here is a simple protocol in a randomized query model that shows that $RCC(MMS) = \Theta(n)$, for identical binary valuations, with $v(\items) = n$. A query is a partition of the set of items into two subsets. The reply specifies for each of these subsets whether it has value~0 (in which case it is discarded), value~1 (in which case it is selected as one of the bundles of the final allocation), or value larger than~1 (in which case one applies the protocol recursively to that subset). It is not hard to see that if the queries are selected uniformly at random (each item of the set is selected with probability $\frac{1}{2}$ of being in each of the parts, independently of other items), the protocol terminates in $O(n)$ queries, in expectation. 

One approach that we use in this paper to design efficient protocols is to use contiguous allocations, when they exist. For binary valuations, contiguous MMS allocations can easily be seen to exist. However, using them would lead to requiring communication complexity $\Omega(n \log \frac{m}{n})$, even if $v(\items) = n$. This is because the $n$ items of value~1 might be arranged in $\frac{n}{2}$ pairs of consecutive items of value~1, and then the $n-1$ break points of a contiguous MMS allocation would need to include the $\frac{n}{2}$ break points within pairs. A hitting set for such allocations would be of size at least $\frac{\binom{m - n/2}{n/2}}{\binom{n}{n/2}} \ge (\frac{m}{n})^{\Omega(n)}$. With not much additional work (and using Lemma~\ref{lem:lowerBounds}), this lower bound on deterministic communication complexity can be extended also to randomized communication complexity, though we omit the details.

The above discussion leads to the following theorem.

\begin{theorem}
    \label{thm:binaryMMSIdentical}
    Consider MMS allocations for instances in which all agents have the same unit demand valuation (or the same binary valuation $v$, with $v(\items) = n$), with $m \ge 2n$.  In this setting, communication complexity equals description complexity (as valuations are identical), and satisfies:
    
    \begin{itemize}
        \item $CC(MMS) = \Theta(n + \log\log m)$.
        \item $RCC(MMS) = \Theta(n)$.
        \item If one insists on contiguous allocations, then $CC(MMS) = \Theta(n \log \frac{m}{n})$.
    \end{itemize} 
\end{theorem}

\subsection{Non-identical unit demand valuations}

We now consider instances in which different agents may have different unit demand valuations (or additive binary valuation $v$, with $v(\items) = n$).

One round of the Round Robin protocol, giving all remaining items to the last agent, show that $CC(MMS) \le (n-1) \log m$. 

We now describe a deterministic protocol that has a better dependence on $m$. Partition $\items$ into $n$ equal size bundles (equal up to one item), and give each agent one of the bundles. Each agent sends one bit, specifying whether she is satisfied (her bundle has value at least~1 for her), or not (her bundle has value~0 for her).
Thereafter, each unsatisfied agent $i$ in her turn replaces her bundle by a bundle that satisfies her, as follows. 

\begin{enumerate}
    \item If there is a different unsatisfied agent $j$ whose bundle satisfies $i$, then:
    \begin{enumerate}
        \item Agent $i$ names such an agent $j$ (taking $\log n$ bits of communication).
        \item Agents $i$ and $j$ swap their bundles with each other.
        \item Agent $j$ sends one bit to specify whether she is satisfies with her new allocation.
    \end{enumerate}  
    
    \item If there is no agent $j$ as above, then there must be some agent $\ell$ whose bundle has value at least~2 for agent $i$. 
    
    \begin{enumerate}
        \item Agent $i$ names such an agent $\ell$ (taking $\log n$ bits of communication).
        \item  Agent $i$ specifies a partition of the bundle of agent $\ell$ into two parts, such that each part has value at least~1 for agent $i$. As the bundle of agent $\ell$ has size at most $\lceil \frac{m}{n} \rceil$, this takes $O(\log \log \frac{m}{n})$ bits (see Section~\ref{sec:unitDemand}). 
        \item Agent $\ell$ gets to keep the part that she prefers (this uses one bit of communication), and additional items from the bundle of agent $i$ so at to keep the size of her bundle unchanged (selecting these additional item does not require communication -- they can be the first items in $i$' bundle). Agent $i$ gets the remaining items from the bundles of $i$ and $\ell$. (Now both agents are satisfied, and each bundle still has size $\frac{m}{n}$, up to one item.) 

    \end{enumerate}
\end{enumerate}

The above proves the following Theorem.

\begin{theorem}
    \label{thm:binaryMMSDeterministic}
    Consider instances in which all agents have (possibly different) unit demand valuations (or binary valuations $v_i$, with $v_i(\items) = n$), with $m \ge 2n$.  In this setting  $CC(MMS) \le O(n (\log n + \log\log \frac{m}{n}))$.
\end{theorem}

We now turn to randomized protocols. We first observe that if randomization is allowed, then step 2(b) of the algorithm above requires only 3 bits in expectation, and not $O(\log \log \frac{m}{n})$ (see Section~\ref{sec:unitDemand}). Thus we immediately have that $RCC(MMS) \le O(n \log n)$, which is independent of $m$. Our goal is to improve this upper bound to $RCC(MMS) \le O(n)$. For this purpose, we may assume that $n$ is larger than some large constant $K$, because for smaller $n$ the term $\log n$ is not larger than the constant $\log K$.

We present now our randomized protocol. It is a randomized version of the deterministic protocol, with an additional modification that introduces a notion of auxiliary bundles. They are introduced so as to simplify the analysis.

{\bf Protocol RUD (Randomized Unit Demand):}

Partition uniformly at random $\items$ into $n$ {\em initial} bundles uniformly at random, giving each agent one of the bundles. Each agent sends one bit, specifying whether she is satisfied (her bundle has value at least~1 for her), or not (her bundle has value~0 for her).
Thereafter, each unsatisfied agent $i$ in her turn replaces her bundle by a bundle that satisfies her. The process for doing so may create up to $n$ auxiliary bundles that are not allocated to any agent. When the algorithm ends, all items in the auxiliary bundles are allocated in an arbitrary fashion (e.g., all of them to agent~1). This last stage does not require any communication. We now describe how an unsatisfied agent $i$ becomes satisfied. This is done by choosing one of the following three types of eligible bundles (there must be an eligible bundle of at least one of these types). The eligible bundle chosen will be the one requiring the least number of bits to communicate the choice, given that the bundles are ordered at random, and bundle names are encoded such that a bundle in location $r$ in this order is encoded using $O(\log r)$ bits.

\begin{enumerate}
\item Every auxiliary bundle that satisfies $i$ is eligible. If selected by agent $i$, then agent $i$ swaps her bundle with this auxiliary bundle. 
    \item A  bundle of an unsatisfied agent $j$ that satisfies agent $i$ is eligible. If selected by agent $i$, then:
    \begin{enumerate}
        \item Agents $i$ and $j$ swap their bundles with each other.
        \item Agent $j$ sends one bit to specify whether she is satisfies with her new allocation.
    \end{enumerate}  
    
    \item A bundle of an agent $\ell$ whose bundle has value at least~2 for agent $i$ is eligible.  If selected by agent $i$, then:
    
    \begin{enumerate}
        \item  Agent $i$ specifies a partition of the bundle of agent $\ell$ into two parts, such that each part has value at least~1 for agent $i$. This requires in expectation at most 2 bits of communication (see Section~\ref{sec:unitDemand}). 
        \item Agent $\ell$ gets to keep the part that she prefers (this uses one bit of communication), and agent $i$ gets the other part of the bundle of agent $\ell$. The previous bundle of agent $i$ is declared to be an auxiliary bundle, not held by any agent.

    \end{enumerate}
\end{enumerate}

Making an agent satisfied requires $O(1)$ bits of communication, plus the number of bits required in order to transmit the identity of the selected eligible bundle. THe following lemma will allow us t upper bound this last quantity.

\begin{lemma}
    \label{lem:BallsInBins}
    With probability at least $\frac{1}{n}$, for sufficiently large $n$, for every agent $i$, for every $k$ in the range $1 \le k \le n$, the smallest set of initial bundles whose sum of values exceeds $k$ contains at least $f(k) = \lceil \frac{k^2}{e^2 n} \rceil$ bundles (where $e$ is the base of the natural logarithm).  
\end{lemma}

\begin{proof}
For an agent $i$, each valuable item (of value~1 to the agent) is placed in a uniformly random initial bundle, independently of other items (similar to the classical {\em balls in bins} setting). The probability that there exists a set of $f(k)$ initial bundles that contains $k$ valuable items is $\binom{n}{k} \cdot \binom{n}{f(k)} \cdot (\frac{f(k)}{n})^k$. To prove the lemma, we use a union bound over the $n$ agents, and all values of $k$. Thus we need to establish that

\begin{equation}
    \label{eq:balls}
    \binom{n}{k} \cdot \binom{n}{f(k)} \cdot \left(\frac{f(k)}{n}\right)^k \le \frac{1}{n^3}
\end{equation}

Observe that for our choice of $f(k) = \lceil \frac{k^2}{e^2 n} \rceil$, one may assume that $k > \sqrt{n}$, because for $k \le \sqrt{n}$ we have that $f(k) < 1$, and it is always the case that at least one initial bundle is needed in order to reach a positive value.

To prove inequality (\ref{eq:balls}) with $f(k) = \lceil \frac{k^2}{e^2 n} \rceil$ and $k > \sqrt{n}$, we note that $f(k) \le \frac{k}{2}$. Thus (\ref{eq:balls}) is implied by:

\begin{equation*}
    \binom{n}{k} \cdot \binom{n}{k/2} \cdot \left(\frac{\lceil \frac{k^2}{e^2 n} \rceil}{n}\right)^k \le \frac{1}{n^3}
\end{equation*}

Using standard approximations, the above inequality simplifies to roughly $\left( \frac{k}{en} \right)^{k} \le \frac{1}{n^6}$. For sufficiently large $n$ and $k > \sqrt{n}$, this last inequality can easily be seen to hold.
\end{proof}

\begin{theorem}
    \label{thm:binaryMMSRandomized}
    Consider instances in which all agents have (possibly different) unit demand valuations, with $m \ge 2n$.  In this setting $RCC(MMS) \le O(n)$. 
\end{theorem} 

\begin{proof}
    We analyse the expected number of bits needed in order for all agents to specify their respective eligible bundles in our randomized protocol RUD. We use Lemma~\ref{lem:BallsInBins}. Recall that the statement of that Lemma fails with probability at most $\frac{1}{n}$. As specifying an eligible bundle takes at most $O(\log n)$ bits (because there are only $O(n)$ bundles to choose from), the failure of the lemma contributes only $\frac{1}{n} \cdot O(n \log n) = O(\log n)$ to the expectation. Hence we may assume that the statement of Lemma~\ref{lem:BallsInBins} holds.

    Consider the turn of an unsatisfied agent at the time when there are $k$ unsatisfied agents. Then each of the $n - k$ satisfied agents holds at least one of the $n$ top items for that agent. Fixing these $n - k$ items (one for each satisfied agent), we still have $k$ remaining top items. Observe that any bundle that contains at least one of these remaining items is an eligible bundle. This implies that the number of eligible bundles is at least $f(k)$ from Lemma~\ref{lem:BallsInBins}, namely, at least $\lceil \frac{k^2}{e^2 n} \rceil$. As there are at most $2n$ bundles altogether and they are ordered at random, specifying the first eligible bundle requires $O(\log \frac{2n}{f(k)})$ bits in expectation, which is $O(\log \frac{n}{k})$.

    It follows the the expected number of bits spent by the RUD protocol in order to specify eligible bundles is $O(\sum_{k=1}^{n} \log \frac{n}{k}) = O(\log \frac{n^n}{n!}) \simeq O(\log (e^n)) = O(n)$. As the expected number of additional bits spent by the RUD protocol is also $O(n)$, the theorem is proved.
\end{proof}

\subsection{Communication versus description complexity}
\label{sec:CversusD}

Recall that we assume that $m \ge 2n$. Here we provide a lower bound for the case that $m$ is very close to $n$, including $m = n+1$.

\begin{lemma}
    \label{lem:lowerSmallm}
    For $m > n$, 
the randomized communication complexity of MMS allocations for unit demand valuations is at least $n-1$. 
\end{lemma}

\begin{proof}
Let $S$ denote the set of first $n+1$ items. Suppose that for each valuation function $v_i$ it is the case that exactly $n$ of the items of $S$ have value~1, and all remaining items have value~0. In every MMS allocation there must be $n-1$ agents that each receive exactly one item from $S$. Every agent $i$ that receives only one item from $S$ must have communicated at least one bit in the allocation protocol, so as to approve that she values the bundle that she receives as having value at least~1 (no other agent can approve on behalf of agent $i$, as no other agent knows $i$'s valuation function). Hence, the total number of bits communicated is at least $n-1$. 
\end{proof}

The next proposition shows that when $m = n+1$, the description complexity is much lower than the (randomized) communication complexity. 

\begin{proposition}
\label{pro:unitDemandDC}
    For instances with unit demand valuations and $m = n+1$ items, $DC(MMS) \le \lceil \log n \rceil$.   
\end{proposition}

\begin{proof}
    Let the items be $\{e_0, \ldots, e_n\}$. To get her MMS, each agent should get one of her top $n$ items. Hence any agent receiving two items gets her MMS. For each $j \in [1, \ldots, n]$, consider the cyclic allocation in which each agent $i \in \{1, \ldots, n-1\}$ gets item $e_{\{i + j \mod n+1\}}$, and agent $n$ gets the remaining two items. At least for one such value of $j$, this is an MMS allocation, as each agent $i \in \{1, \ldots, n-1\}$ excludes only one value of $j$ (for which $e_{\{i + j \mod n+1\}}$ is her bottom item). Specifying this value of $j$ requires only $\lceil \log n \rceil$ bits.
\end{proof}

\subsection{EF1 for unit demand}

With unit demand valuations, the following simple deterministic protocol produces and EF1 allocation. Querying the agents in an arbitrary order, each agent selects in her turn the item most valuable to her among those remaining. The last agent receives all remaining items. The communication complexity of this protocol is $(n-1)\cdot \lceil \log m \rceil$. We show that randomized protocols can to better.

\begin{proposition}
\label{pro:unitDemandEF1}
    For instances with unit demand valuations, 
    $RCC(EF1) \le O(n \log n)$.   
\end{proposition}

\begin{proof}
We may assume that $m > n^3$, as otherwise the deterministic protocol has communication complexity $O(n\log n)$. 

Partition $\items$ uniformly at random to $n^3$ bundles. Such a partition is {\em successful} if for every agent, every one of her $n$ top items is in a different bundle. The probability that the partition fails to be successful is at most $n \cdot \binom{n}{2}\cdot n^3 \cdot (\frac{1}{n^3})^2 \le \frac{1}{2}$ (by taking a union bound over the events that one bundle has two specific items out of the $n$ top items of one agent). Every agent is asked whether from her point of view the partition is successful, and provides a one bit answer. If at least one agent declares that it fails, then a new random partition is chosen. This process is repeated until a successful partition is selected. In expectation, this happens on the second attempt, so the expected number of bits communicated until a successful partition is reached is $O(n)$.

Once a successful partition is reached, we view every bundle of it as a single new item in a new instance with $n^3$ items. The value of such a new item to a unit demand bidder is the maximum among the values of the items in the bundle that corresponds to the new item. Running the simple deterministic protocol on the new instance gives an EF1 allocation for the new instance. This is also and EF1 allocation for the original instance, because for every agent, the top $n$ new items have the same values as the top $n$ items in the original instance. 
\end{proof}

We now prove a lower bound matching the upper bound of Proposition~\ref{pro:unitDemandEF1}.

\begin{theorem}
    \label{thm:EF1}
    For $m \ge 2n$ and agents with unit demand valuations, the randomized description complexity of EF1 allocation is $\Omega(n\log n)$. 
\end{theorem}

\begin{proof}
    We may assume that $n > n_0$ for some sufficiently large constant $n_0$, by adjusting the constant in the $\Theta(n\log n)$ notation. We may also assume that $m = 2n$, because if $m > 2n$, we can restrict attention to valuations in which the top $n$ items are within the set $\{e_1, \ldots, e_{2n}\}$.
    
    We consider the class $V_{m,k}$ of binary valuation functions over $m$ items, in which $k$ items have value~1, and $m - k$ items have value~0. (The items of value~1 represent the top $k$ items in a unit demand valuation. In this unit demand valuation, the $k+1$st most valuable item has value strictly smaller than the $k$th most valuable item.)
    We fix $k$ to have value roughly $m^{3/4}$. Note that if an agent $i$ with $v_i \in V_{m,k}$ receives a bundle that contains only small items, and some other agent receives a bundle that contains at least two large items, then the allocation is not EF1. Based on $V_{m,k}$, we consider the class $C_{m,k}$ of allocation instances in which each agent has a valuation function from $V_{m,k}$. Crucially for our proof, different agents may have different valuation functions.

    Consider an arbitrary allocation $A = (A_1, \ldots, A_n)$. We compute an upper bound on the probability that $A$ is an EF1 allocation, when the input instance is chosen at random from $C_{m,k}$. Without loss of generality, no bundle $A_i$ is empty, as otherwise $A$ cannot be EF1 (we may assume that the original unit demand valuations has at least $n$ items of positive value). This, combined with the requirement $m = 2n$, implies that at least half the bundles each has no more than~2 items. Consider an agent $i$ that received a bundle $A_i$ with $|A_i| \le 2$. Over the choice of random instance from $C_{m,k}$, $v_i(B_i) = 0$ with probability at least $1 - \frac{2k}{m}$. If indeed $v_i(B_i) = 0$, then for $A$ to be EF1, there should not be any bundle $A_j$ with $v_i(A_j) \ge 2$. For the $m - 2$ items not in $A_i$, partition them into disjoint pairs of items and possibly some leftover items (that cannot find a feasible partner), such that for each pair, both items of the pair are in the same bundle. This gives at least $\frac{n-1}{2}$ distinct pairs. (In every bundle that contains an odd number of items, one item is wasted. Hence at least $m-2 - (n-1) = n-1$ items are in pairs.) The probability that none of these pairs is composed of two items that have value~1 under the random $v_i$ is at most $(1 - {\frac{k(k-1)}{(m-2)(m-3)}})^{\frac{n-1}{2}}$ (as the events are negatively correlated). This probability is smaller than $\frac{k}{m}$ for our choice of $k$ and $n > n_0$. Hence the probability that the allocation is EF1 with respect to this agent $i$ is at most $\frac{k}{n}$. 

    As there are at least $\frac{n}{2}$ agents that each receive at most two items, and their valuation functions are chosen independently when the allocation instance is chosen at random from $E_{m,k}$, the probability that $A$ is an EF1 allocation is at most $(\frac{k}{n})^{\frac{n}{2}} = n^{-\Omega(n)}$. The lower bound of $\Omega(n \log n)$ on the communication complexity follows from Lemma~\ref{lem:lowerBounds}.
\end{proof}


\section{Binary valuations}

We first prove a lower bound on the randomized communication complexity of MMS allocations for binary valuations. The lower bound holds even if all agents have the same valuation functions, and hence is also a lower bound on randomized description complexity.

\begin{theorem}
    \label{thm:binaryLower}
    For $m \ge 2n$, the randomized description complexity of MMS allocations satisfies $RDC(MMS) \ge \Omega(n\log(\frac{m}{n}))$, even if all agents have the same binary valuation function. 
\end{theorem}

\begin{proof}
Let $k$ be the largest integer for which $m \ge 2kn$. Note that $k \ge 1$, by the requirement that $m \ge 2n$.
We say that a binary valuation function $v$ is {\em balanced} if $v(\items) = kn$. Observe that if all agents have the same balanced valuation function $v$, then in every MMS allocation $A = (A_1, A_2, \ldots A_n)$, it must be the case that $v(A_i) = k$, for all $i$. 
 Consider the set $C_{n,m}$ of allocation instances in which all agents have the same binary valuation function $v$, and $v$ is balanced.

Given an allocation $A$, we analyse the probability that it {is} MMS for an input instance chosen uniformly at random from $C_{n,m}$.
 To upper bound this probability, it is convenient to consider a distribution $D$ over binary valuations in which the value of each item is sampled independently at random. Each item independently has value~1 with probability $p = \frac{kn}{m}$, and value~0 otherwise. Note that $\frac{1}{4} < p \le \frac{1}{2}$. For a valuation $v$ sampled at random from $D$, in expectation, $v(\items) = kn$. Consequently, the probability that a valuation function sample from $D$ is balanced is $\Omega(\frac{1}{\sqrt{m}})$. Moreover, conditioned on being balanced, each balanced valuation is equally likely to be sampled, and hence we get a uniformly random sample of input instance from $C_{n,m}$.

 Consider an arbitrary allocation $A = (A_1, A_2, \ldots A_n)$. For $v$ selected uniformly at random from $D$ we have that $Pr[v(A_i) = k] \le \min[\frac{1}{2}, O(\frac{1}{\sqrt{k}})]$ for every $A_i$, because $\frac{1}{4} < p \le \frac{1}{2}$.  As the events $v(A_i) = k$ are independent for different values of $i$, the probability that $v(A_i) = k$ simultaneously for all $i$ is $(\min[\frac{1}{2}, O(\frac{1}{\sqrt{k}})])^n$. 
 Using the fact that $k = \Theta(\frac{m}{n})$, this probability can be seen to be of the form $(\frac{n}{m})^{\Omega(n)}$. The same holds for input instances sampled uniformly at random from $C_{n,m}$, because sampling from $C_{n,m}$ instead of $D$ can increase the probability by at most $O(\sqrt{m})$, which only affects the constants hidden in the $\Omega$ notation in our bounds (details omitted).
 
 Applying Lemma~\ref{lem:lowerBounds}, we get that $RDC(MMS) \ge \Omega(n \log \frac{m}{n})$.
\end{proof}

Perhaps surprisingly, when agents may have different binary valuations, the randomized communication complexity for MMS allocations is only $O(n\log(\frac{m}{n}))$. This matches (within constant multiplicative factors) the lower bound in Theorem~\ref{thm:binaryLower}, which holds even if all agents have the same binary valuation function. We remark that a bound of $RCC(MMS,Binary) \le O(n \log m)$ is implied by {Proposition~\ref{pro:randProp1}}, because for binary valuations, Prop1 allocations coincide with MMS allocations. Hence the main content of the following theorem is replacing $\log m$ by $\log \frac{m}{n}$, which is very significant in the special case that $m = O(n)$.

\begin{theorem}
\label{thm:binaryUpper}
    If all agents have binary valuation functions, then the randomized communication complexity for finding an MMS allocations is $O(n\log \frac{m}{n})$.
\end{theorem}

\begin{proof}
We assume  that $m = kn$ for some integer $k \ge 2$. The assumption can be made without loss of generality because we always assume that $m \ge 2n$, and one can add at most $n-1$ dummy items of value~0 to $\items$ so that $m$ becomes a multiple of $n$. We may further assume that {$k \le n^{\frac{1}{15}}$}, as otherwise the theorem is implied by {Proposition~\ref{pro:randProp1}}. 
{We may also assume that $n > n_0$ for some sufficiently large constant $n_0$, as otherwise the communication complexity is $O(1)$ (as we assume that $m < n^2$).}

{For each agent $i$ there is an integer $k_i$ for which the binary valuation function $v_i$ is such that the number of {\em large} items (those of value~1) is at least $k_i \cdot n$ and at most $(k_i + 1)\cdot n - 1$. The remaining items are {\em small}, of value~0. In an MMS allocation, agent $i$ needs to receive at least $k_i$ large items. Without loss of generality, we may assume that $k_i > 0$, as otherwise agent $i$ places no constraints on MMS allocations.} 
Observe that $\max_{i \in [n]} k_i \le k$.  

In our proof we assume that items are arranged in a uniformly random order $\pi$. For this reason, our upper bounds refer to random communication complexity.

Our algorithm has three phases.

In the first phase, each agent $i$ reports her respective $k_i$ value. This requires $O(\log k)$ bits per agent, contributing $O(n\log k)$ to the communication complexity. Using these reports, agents are sorted from smallest $k_i$ value to largest, using an arbitrary tie breaking rule. We rename agents so that now we have $k_i \le k_{i+1}$ for all $i \in \{1, \ldots, n-1\}$.

In the second phase, each agent $i$ in her turn, starting from agent~1 and continuing until we either run out of agents or run out of items, selects a minimal prefix of the remaining items that gives her a value of $k_i$. That is, agent~1 reports the minimal location $j_1$ such that $v_1(\{e_1, \ldots, e_{j_1}\}) = k_1$. Then, agent~2 reports the minimal location $j_2$ such that $v_1(\{e_{j_1+1}, \ldots, e_{j_2}\}) = k_2$, and so on. If all agents manage to provide such a report before we run out of items, then we have an MMS allocation, and we are done. Hence, we assume here that for some $\ell > 0$, only the first $n-\ell$ agents supply such a report, whereas agent $n-\ell+1$ failed to do so. At the end of the second phase we have $n-\ell$ {\em happy} agents that have an MMS bundle, and $\ell$ {\em sad} agents that have not yet received any item. The set $\items$ is partitioned into $n-\ell$ {\em main sets} that go to the happy agents, and a set of leftover items (this last set might be empty). We let $t$ denote the size of the largest main set. 

The communication complexity of the second phase is $O(n \log k)$, as the second phase involves specifying at most $n$ items out of $m= nk$, in increasing order. This is equivalent to specifying lengths of at most $n$ intervals, whose sum of lengths is at most $kn = m$. Specifying the length $x$ of an interval takes $O(\log x)$ bits (see Section~\ref{sec:preliminaries}). 

The third phase has $\ell$ rounds. In round $r$ agent $n-\ell+r$ selects large items until she reaches her desired value of $k_{n-\ell+{r}}$. These items may be taken either from the set of leftover items, or from the main sets. Selecting an item from a main set $j$ might cause happy agent $j$ to become unhappy. We now explain how to implement each round while avoiding such a problem.

Consider agent $i$ with $i > n - \ell$. For $j < i$, we say that agent $j$ is {\em tight} in round $i$ if agent $j$ holds exactly $k_j$ items (in which case all of them are large for $v_j$). Only agents that were original happy can be not tight. As will be apparent from the next paragraph, non-tight agents hold their original main set, 

If the set of leftover items contains large items  for $v_i$, then agent $i$ selects them (this uses at most $\log m$ bits of communication {per item}). We now explain what to do when the set of leftover items runs out of large items for $v_i$. For this, we note that each agent $j < i$ that is tight holds $k_j \le k_i$ items (this is the place in our proof in which we use the first phase, that implies that agents are sorted so that the sequence of $k_i$ values is non-decreasing). The total number of agents other than $i$ is only $n-1$, agent $i$ so far has fewer than $k_i$ items, and the leftover set does not contain any large item that is large for $v_i$. Consequently, it must be that the main set of at least one of the non-tight agents $j$ contains at least $k_i + 1$ items that are large for $v_i$. Agent $i$ first points to this agent $j$ (using $\log n$ bits). Then agent $j$ selects exactly $k_j$ large items from her main set (using at most $k \log t$ bits), and releases the remaining items so that they are moved into the set of leftover items. This makes agent $j$ tight. As for agent $i$, she can select at least one of these released items, as $k_i + 1 > k_j$. Specifying this item requires at most $\log t$ bits.

The communication complexity of the third phase can be upper bounded as follows. There are $\ell$ agents that select items in this phase. Each agent selects at most $k$ items. For each item selected, if it is taken from the leftover set, the communication complexity is $\log m$. If it is taken from a main set of an agent $j$, then the communication complexity is at most $\log n + (k+1)\log t$, due to agent $i$ naming agent $j$, agent $j$ selecting $k_j$ large items from her main set, and agent $i$ selecting an item from the released items. 
Hence overall, the communication complexity of the third phase is at most $\ell \cdot k \cdot \max[\log n + (k+1)\log t, \log m] \le O(\ell \cdot k^2 \cdot \log m)$. 

We now bound the randomized communication complexity. For the first two phases, the deterministic complexity is within our bounds of $O(n \log k)$. Hence it remains to bound the expected communication complexity of the third phase. For this, we need to to bound $E[\ell]$, the expectation of $\ell$. Though we believe that $E[\ell] \simeq O(\sqrt{n})$, in Lemma~\ref{lem:binary} (see Section~\ref{app:binary}) we provide a weaker bound of $E[\ell] \le 4n^{\frac{6}{7}}$, that is easier (for us) to prove, and suffices for out purpose. (The proof of Lemma~\ref{lem:binary} is based on the fact that $\pi$ is a random ordering. The aspect that complicates its proof, and motivates us to settle for a sub-optimal upper bound on $E[\ell]$, is that {even when $\pi$ is random, the sizes of the main sets are not independent from each other}.) Using this bound, we deduce that the expected communication complexity of the third phase is $O(n^{\frac{6}{7}} \cdot k^2 \cdot \log m) < n$ {(the inequality holds when $n$ is sufficiently large, due to the assumption that $k \le n^{\frac{1}{15}}$)}, which is negligible compared to communication complexity of the first two phases.
\end{proof}


\section{Valuations supported on two values $a > b \ge 0$}

In 2-valued valuations, there are two prespecified values $a > b \ge 0$, and all valuations are additive, where every single item has either value $a$ or $b$. Binary valuations are a special case of 2-valued valuations, in which $b = 0$. We show that change $b$ to be slightly larger than~0 makes a dramatic difference to the communication complexity of MMS allocations.

\begin{theorem}
    \label{thm:2valuedLower}
    For $n \ge 2$ agents and $m \ge 2n$ items, the randomized description complexity of MMS allocations is $\Omega(m)$, even if all agents have the same 2-valued valuation.
\end{theorem}

\begin{proof}
Proving a lower bound of $\Omega(n)$ is relatively straightforward (when $m \ge 2n$), and is omitted. Hence we may assume that $m \ge 4n$. Let integer $k\ge 2$ be such that $2kn < m \le 2kn + 2n$. 
We consider a class of allocation instances that we denote by $C_{n,k,m}$. In every instance $I \in C_{n,k,m}$, there are $n$ agents and they all have the same valuation function $v_I$. Under $v_I$, there are $m' = kn + n - 1$ {\em large} items with value~1, and $m - m'$ {\em small} items with 
value $\frac{1}{m-m'}$.
There are $\binom{m}{m'}$ possible choices for $v_I$, and so $|C_{n,k,m}| = \binom{m}{m'}$, which is $2^{\Theta(m)}$.

For every allocation instance $I \in C_{n,k,m}$, in every MMS allocation, $n-1$ agents receive $k+1$ large items, and one agent receives $k$ large items and all the small items.


Consider an arbitrary allocation $A_1, \ldots, A_n$ that is claimed to be MMS. For instances in $C_{n,k,m}$ all agents have the same valuation function. Hence, we may assume without loss of generality that $A_n$ has the largest number of items. Moreover, it must be that $|A_i| = k+1$ for all $1 \le i \le n-1$, and $|A_n| = m - m' + k$, as this is true for all MMS allocations for instances in $C_{n,k,m}$. Pick an allocation instance $I$ uniformly at random from $C_{n,k,m}$. For $A_1, \ldots, A_n$ to be an MMS allocation for $I$, it must be that all items in $A_1, \ldots, A_{n-1}$ are heavy. This event has probability $\frac{\binom{m'}{(n-1)(k+1)}}{\binom{m}{(n-1)(k+1)}} = 2^{-\Theta(m)}$. Lemma~\ref{lem:lowerBounds} now implies the $\Omega(m)$ lower bound on randomized description complexity.
\end{proof}

We prove an upper bound on communication complexity that matches up to a multiplicative factor of $O(\log n)$ the lower bound of Theorem~\ref{thm:2valuedLower}. 

\begin{theorem}
    \label{thm:2valued}
    For instances with 2-valued additive valuations, $CC(MMS) \le O(m \log n)$. In fact, there is a deterministic protocol in which each agent sends $O(\frac{m}{n} \log n)$ bits.
\end{theorem}

\begin{proof}
    Let the two possible values of items be $a > b \ge 0$. For each agent $i$, let $m_i$ denote the number of items of value $a$ (and $m - m_i$ is the number of items of value $b$). Each agent $i$ reports her $m_i$ value, using $\log m$ bits.

    Given the $m_i$ values, the algorithm assumes that the instance is {\em ordered}. Namely, that for every agent $i$, the items of value $a$ are $\{e_1, \ldots, e_{n_i}\}$, and the remaining items have value $b$. Under this assumption, the algorithm produces an MMS allocation, using for example the algorithm of~\cite{feige22}. In this allocation, for every agent $i$ there is a number $a_i$ of items of value $a$ that she receives and a number $b_i$ of items of type $b$ that she receives, such that $v_i(a_i \cdot a + b_i \cdot b)$ is at least the MMS of agent $i$. If $a_i > \frac{m}{n}$, we replace $a_i$ by $\lfloor \frac{m}{n} \rfloor < a_i$, and replace $b_i$ by $b_i + a_i - \lfloor \frac{m}{n} \rfloor$, and still maintain that agent $i$ gets at least her MMS, and that the total number of items allocated is $m$. These values of $a_i$ and $b_i$ for each agent $i$ are declared. (This requires no communication by the agents.)

    Order the agents by their $m_i$ values, from smallest to largest. Hence we have that $m_1 \leq m_2 \ldots \leq m_n$. As the algorithm computed an allocation for the ordered instance, it holds that for every $k$, $m_k \ge \sum_{i=1}^k a_i$. 

    Scan the agents one by one. Each agent $i$ in her turn picks among the remaining items $a_i$ items of value $a$ for her. By the condition that $a_i < \frac{m}{n}$, this takes at most $\log \binom{m}{m/n} = O(\frac{m}{n} \log n)$ bits per agent.

    Finally, each agent $i$ gets $b_i$ arbitrary additional items (this requires no communication by the agents).

The number of bits communicated by each agent is $O(\log m + \frac{m}{n} \log n) = O(\frac{m}{n} \log n)$, as desired.
\end{proof}



\section{Additive valuations}

In this section, we prove the results that concern additive valuations in general: Theorem~\ref{thm:MMSadditive} and Theorem~\ref{thm:Aprop}.

We remark that the lower bound for MXS allocations, Corollary~\ref{cor:binaryMXS}, was already implicitly proved in Theorem~\ref{thm:2valuedLower}. This is because for the family of allocation instances considered in the proof of Theorem~\ref{thm:2valuedLower}, every MXS allocation is also an MMS allocation.

\subsection{MMS allocations}

Here we prove Theorem~\ref{thm:MMSadditive}. By the discussion following that theorem, it suffices to prove he following proposition.

\begin{proposition}
    \label{pro:MMSidetical}
    When all agents have the same additive valuation function and $m \ge 2n$, then $RDC(MMS) = \Omega(m \log n)$.
\end{proposition}

\begin{proof}
    Let $k$ be the largest integer so that $m \ge kn$. Let $K$ be a sufficiently large integer ($K \ge m^2$ suffices). Consider valuation function $v$ such that for every $1 \le j \le n$, there are $k-1$ {\em medium} items of value $K - j$, and one {\em large} item of value $K^2 + (k-1)j$. The MMS is $K^2 + (k-1)K$, because for every $1 \le j \le n$ we can create a bundle containing $k-1$ items of value $K - j$ and one item of value $K^2 + (k-1)j$. Moreover, if $K$ is sufficiently large then the above is the only MMS partition (the bundle containing the item of value $K^2 + k-1$ must also contain the $k-1$ items of value $K - 1$, and so on).  

    If item names are random, then any allocation has probability at most $\frac{((k-1)!)^n}{((k-1)n)!}\simeq (\frac{1}{n})^{(k-1)n} \le n^{-\frac{m}{2}}$ for being an MMS allocation. (The upper bound on the probability was computed by first revealing which are the large items and which are the medium items, and then also revealing the values of the large items. There are $((k-1)n)!$ possible permutations over the medium items, of them $((k-1)!)^n$ match the medium items correctly to their respective large items.) The proof of the proposition now follows from Lemma~\ref{lem:lowerBounds}.
\end{proof}

\subsection{Prop1 allocations}
\label{sec:Prop1}

As a warm up towards proving Theorem~\ref{thm:Aprop}, in this section we consider the easier task of producing Prop1 allocations, dropping the requirement that allocations are also $\frac{n}{2n-1}$-TPS. 

First, we consider the lower bound of $\Omega(n \log \frac{m}{n})$, that as stated in Theorem~\ref{thm:Aprop}, applies even to Prop1 allocations. This lower bound is an immediate consequence of Theorem~\ref{thm:binaryLower}, because for binary valuations there is no difference between MMS allocations and Prop1 allocations, under our definition~\ref{def:Prop1} of Prop1.

\begin{remark}
\label{rem:lowerProp1}
    With a little extra work, the lower bound in Theorem~\ref{thm:binaryLower} extends also to the weaker definition of Prop1 in which one additional item brings the agent to a value at least as large as her proportional share (rather than strictly larger). In the proof, replace the balance condition of $v(\items) = kn$ by $v(\items) = kn+1$. Further details are omitted.
\end{remark}

We now prove upper bounds on the communication complexity.
Procedures for fair allocation of a divisible good 
can sometimes be adapted to produce fair allocations for indivisible goods. In particular, this was used in~\cite{Suksompong19} to show that there always is a so called {\em contiguous} Prop1 allocation, a fact that we shall make use of in establishing communication complexity upper bounds. Specifically, procedures of~\cite{EP84} for allocation of a divisible good using few cuts easily translate to low communication allocation procedures that produce contiguous Prop1 allocations. The relatively simple Prop1 case will serve as a starting point for our more advanced results, that will require us to depart from contiguous allocations.

\begin{proposition}
    \label{pro:detProp1}
    The deterministic communication complexity for finding Prop1 allocations satisfies $CC(Prop1) \le n \log m \log n$.
\end{proposition}

\begin{proof}
View the items as being layed out from left to right in the order $e_1$ to $e_m$. As proved in~\cite{Suksompong19}, for every agent $i$, there is a partition $A^i = (A_1^i, \ldots, A_n^i)$ of the items into $n$ consecutive bundles that each is a Prop1 bundle with respect to $v_i$. 

    The protocol for producing a Prop1 allocation is is taken from~\cite{EP84}, and proceeds by binary search. The proposed protocol can easily be adjusted to handle all values of $n$, but for simplicity of notation (so as to avoid the use of ``floor" and ``ceiling" notation), we assume that $n$ is a power of~2.  The protocol works in $\log n$ phases. In phase~1, each agent $i$ sends a message of length $\log m$, specifying which item is the right-most item in her $A_{\frac{n}{2}^i}$ bundle. We refer to these items as {\em cut items}. Sort the cut items from left to right (breaking ties in favor of lowered named agents), and select the item at location $\frac{n}{2}$ in this sorted order (we refer to it as the {\em median}). Using the median, the problem decomposes into two sub-problems. The $\frac{n}{2}$ agents with lowest cut items continue with the left part of the items, and the other agents with the right part.

    In phase~2, each of the left and right parts is decomposed into two sub-problems independently, by employing the median procedure on the respective part (with $\frac{n}{2}$ agents instead of $n$ agents). This decomposition continues recursively until each part contains a single agent. 

    When the protocol ends, one gets a Prop1 allocation of $\items$, as each agents $i$ gets a bundle $B_i$ that contains one of her Prop1 bundles $A_J^i$.

    Each agent participates in $\log n$ rounds and sends $\log m$ bits per round, for total communication complexity of $n \log m \log n$. (In fact, the communication complexity is somewhat lower, because as rounds progress the number of items per-part decreases. However, this saving is not sufficiently interesting to justify discussing it further here.)  
\end{proof}

A randomized procedure of~\cite{EP84} for proportional division of a divisible good naturally translates to give a randomized Prop1 allocation procedure.

\begin{proposition}
    \label{pro:randProp1}
    The randomized communication complexity for finding Prop1 allocations satisfies $RCC(Prop1) \le O(n \log m)$. 
\end{proposition}

The proof of Proposition~\ref{pro:randProp1} is a variation on the proof of Proposition~\ref{pro:detProp1}. The main difference is that a randomized protocol is used in order to find the median cut item, and its expected communication complexity is smaller than that of the deterministic protocol used in the proof of Proposition~\ref{pro:detProp1}. The full proof appears in Section~\ref{app:Prop1}.

\subsection{$\frac{n}{2n-1}$-TPS allocations}
\label{sec:TPS}

We now consider the complementary task of producing $\frac{n}{2n-1}$-TPS allocations, dropping the requirement that allocations are also Prop1. Recall that for Prop1 allocations
there are contiguous allocations. For $\frac{n}{2n-1}$-TPS allocations, this is no longer true. Suppose for example that $n$ is divisible by~4, there are $m = \frac{7n}{4}$ items, all agents have the same valuation function, with $\frac{3{n}}{4}$ ``large" items each having value~1, and $n$ ``small" items each having value $\frac{1}{4}$. In a Prop1 allocation, it suffices that each agent gets at least one item. However, in a $\frac{n}{2n-1}$-TPS allocation, each agent must get value strictly greater than $\frac{1}{2}$, meaning that she must either get at least one large item, or at least three small items. If there is no index $j \in \{1, \ldots, m-2\}$ such that the three items $\{e_j, e_{j+1}, e_{j+2}\}$ are small, then in every contiguous allocations, at most $\frac{3n}{4}$ of the pieces have value strictly above $\frac{1}{2}$, and consequently, $\frac{n}{4}$ of the agents fail to get an $\frac{n}{2n-1}$-TPS allocation.

To design protocols that produce $\frac{n}{2n-1}$-TPS allocations, we adapt a simple algorithm of~\cite{garg2019approximating} to our purpose. That algorithm was shown in~\cite{garg2019approximating} to give a $\frac{1}{2}$-MMS allocation, but a straightforward modification of it gives a $\frac{n}{2n-1}$-TPS allocation. In our context, this translates to an allocation algorithm that includes two phases.

In the first phase, agents are visited in a sequential fashion. When agent $i$ is visited, then if among the remaining items there is an item of value at least  $\frac{n}{2n-1}$-TPS for agent $i$, then agent $i$ selects such an item and leaves. If there is no such item, agent $i$ is moved to the second phase. The first phase requires communication of at most $\log m$ bits per agent.

For the second phase, it is not difficult to show that for the remaining agents and items, a contiguous $\frac{n}{2n-1}$-TPS allocation must exist. Hence, we may adapt the procedures of~\cite{EP84} to handle the remaining agents, in a way similar to that done for Prop1 allocations. This shows that CC($\frac{n}{2n-1}$-TPS) $\le O(n \log m \log n)$ and that RCC($\frac{n}{2n-1}$-TPS) $\le O(n \log m)$. Further details are omitted, as stronger results are presented in Section~\ref{sec:Aprop}. Moreover, in Section~\ref{app:randomBundling} we even prove a stronger bound of RCC($\frac{n}{2n-1}$-TPS) $\le O(n \log n)$.

\subsection{Aprop allocations}
\label{sec:Aprop}

In this section, we design protocols that produce Aprop allocations namely, allocations that are both Prop1 and $\frac{n}{2n-1}$-TPS. 
{For this purpose, we consider {\em semi-contiguous} allocations.} 
Like contiguous allocations, a semi-contiguous allocation partitions the sequence $\{e_1, \ldots, e_m\}$ of items into $n$ disjoint sub-sequences of consecutive items, referred to as {\em blocks}. However, some of the blocks might be empty. In addition, one can designate up to $n$ items as {\em holes}. For any block, its associated {\em net-block} is the set of those items that are in the block but not in any hole. In a semi-continuous allocation, each agent gets a distinct net-block, and possibly also one hole (and all holes need to be allocated). If no blocks are empty and there are no holes, then a semi-contiguous allocation is a contiguous allocation. {The $\frac{n}{2n-1}$-TPS allocations of Section~\ref{sec:TPS} are semi-contiguous, but are not Prop1.}

We shall establish that when agents have additive valuations, a semi-contiguous Aprop allocation exists. This implies that the description complexity for Aprop allocations is $O(n \log m)$. Moreover, we design low communication protocols that find such allocations.

\begin{theorem}
\label{thm:Aprop1}
    If agents have additive valuations, then a semi-contiguous Aprop allocation exists. Moreover, the deterministic communication complexity for finding Aprop allocations is at most $(1 + o(1))n \log m \log n$, and the randomized communication complexity and the description complexity are both $O(n\log m)$.
\end{theorem}

{The proof of this theorem appears in Section~\ref{app:Aprop}.}

\section{Discussion and open questions}
\label{sec:open}

In our work we presented randomized protocols for finding fair allocations. In many of our randomized protocols, it suffices that there is a random order over the items, or a random order over the agents, and then all remaining steps of the protocol can be deterministic. Hence one may also view these protocols as deterministic protocols that are analysed in a semi-random model, in which input instances are generated in two steps. First an adversary selects a worst case allocation instance, and then the order of items (or of agents) is permuted at random. Our analysis shows that some deterministic protocols have very low communication complexity when inputs are generated using this semi-random model.

Our work leaves some questions open. One class of such open questions is quite obvious: to close the gaps between upper bounds and lower bounds in those cases that we did not manage to do so. Here we list open questions of a more general nature.

\subsection{Communication versus description complexity}
\label{sec:openCCDC}

In this work we prove many lower bounds on randomized description complexity, and use them as lower bounds for randomized communication complexity. In many of our theorems, these lower bounds match (up to constant factors) our upper bounds on randomized communication complexity. 
The following proposition (proved in Section~\ref{sec:CversusD}) presents a setting in which there is a large gap between communication and description complexity.

\begin{proposition}
    \label{pro:UDgap}
    For unit demand valuations, if $m = n+1$ then $RCC(MMS,UD) = \Theta(n)$, but $DC(MMS,UD) =\Theta(\log n)$.  
\end{proposition}

In Proposition~\ref{pro:UDgap}, the gap is due to the fact that the description complexity is very small. We view it as a major open problem to develop techniques for proving lower bounds on the communication complexity that are higher than $m \log n$, which is the maximum possible value of description complexity. 
Specifically, is the communication complexity of computing MXS allocation for additive valuations larger than $m \log n$? 

\subsection{Does randomization help?}

In many cases, our deterministic allocation protocols have higher communication complexity than the expected communication complexity of our randomized protocols. We suspect that such a gap is sometimes necessary, but could prove it only in a very special case, that of MMS allocations for a constant number of agents with identical Unit Demand valuations (see tables~\ref{tab:MMSdet} and~\ref{tab:MMS}). Can one prove gaps in additional settings?

\subsection{The communication complexity of approximate MMS} 

{For additive valuations,
for some values of $\rho < 1$, $\rho$-MMS allocations (in which each agent gets at least a $\rho$ fraction of her MMS) are feasible. How does the communication complexity vary as a function of $\rho$? For $\rho \le \frac{n}{2n-1}$, Theorem~\ref{thm:Aprop} implies that RCC($\rho$-MMS)$\le O(n \log m)$, because the TPS is at least as large as the MMS. Moreover, in Theorem~\ref{thm:TPSn} we eliminate the dependence of the upper bound on $m$, providing an upper bound of $O(n \log n)$.
We have not presented lower bounds on  RCC($\rho$-TPS), so we leave open the question of whether RCC($\rho$-MMS)$\le O(n)$, with each agent sending only $O(1)$ bits on average.}

Is there some value of $\rho$ (larger than $\frac{n}{2n-1}$) for which $\rho$-MMS allocations are feasible, but require randomized communication complexity significantly larger than $n \log n$, say $\Omega(n^2)$? If so, then this will show a separation result for the communication complexity of achieving different values of $\rho$. 

\subsection*{Acknowledgements}

This research was supported in part by the Israel Science Foundation (grant No. 1122/22).


\newpage

\bibliographystyle{alpha}


\newcommand{\etalchar}[1]{$^{#1}$}

\newpage

\begin{appendix}

    


\section{Some comments on fairness notions}
\label{app:fair}

We present here a few comments on the fairness notions considered in this paper. Throughout this section, we assume additive valuations.

Referring to Prop1 as a share based fairness notion involves a relaxed interpretation of the notion of a share. According to some definitions~\cite{BF22}, shares are required to enjoy some monotonicity properties that Prop1 does not satisfy (increasing the value of an item, the value of Prop1 might decrease). Consequently, Aprop also does not qualify as a share under such definitions. 

The TPS is at least as large as the MMS. Hence our notion of Aprop, which implies $\frac{n}{2n-1}$-TPS, implies also $\frac{n}{2n-1}$-MMS. 

The definition of the TPS can be motivated as follows. It modifies the definition of Prop only if there are over-proportional items. Suppose that $e$ is an over-proportional item for agent $i$. If $i$ receives $e$, then she gets more than her proportional share, and does not demand additional items. If instead $e$ is given to some other agent $j$, then, in the eyes of $i$, agent $j$ is not entitled to additional items, and can be discarded.
We remain with an instance with one less item and one less agent. In this new instance, if no more over-proportional items remain, agent $i$ expects to get her proportional share. But as some over-proportional items might remain, then in this new instance agent $i$ expects to get her TPS, which leads to the recursive definition of the TPS. 

EF1 allocations are Prop1.  
However, they need not provide any good approximation to the MMS, and hence also not for the TPS. For example, if all agents have the same valuation function, with $n-1$ {\em large} items of value~$n$ and $n$ {\em small} items of value~1, then the MMS is $n$, whereas an EF1 allocation may give each of $n-1$ agents one large item and one small item, and the remaining agent only a small item. This is only a $\frac{1}{n}$ fraction of her MMS. Nevertheless, there is a specific algorithm that produces EF1 allocations (not only for additive valuations, but for all monotone valuations), that of~\cite{LiptonMMS04}, that can easily be adapted to in addition give $\frac{n}{2n-1}$-TPS allocations, and hence it produces Aprop allocations. (The adaptation is to give agents their most preferred item among those remaining, when it is their turn to receive an item. For non-additive valuations, the notion of most preferred items is not well defined, but for additive valuations, it is.)

Finally, we show that MXS implies Aprop.

\begin{proposition}
    \label{prop:MXSAprop}
    For additive valuations, every MXS allocation is also an Aprop allocation.
\end{proposition}

\begin{proof}
For simplicity of the presentation, we prove that every EFX allocation is Aprop. The proof that MXS implies Aprop is similar. 

An EFX allocation $(A_1, \ldots, A_n)$ is Prop1 because if $v_i(A_i) < Prop(v_i,\items,\frac{1}{n})$, there must be another bundle $A_j$ with $v_i(A_j) > Prop(v_i,\items,\frac{1}{n})$. By the EFX property, moving any item $e \in A_j$ to $A_i$ gives $v_i(A_i \cup \{e\}) \ge v_i((A_j \setminus \{e\} )\cup \{e\}) = v_i(A_j) > Prop(v_i,\items,\frac{1}{n})$. 

To see that an EFX allocation $(A_1, \ldots, A_n)$ is $\frac{n}{2n-1}$-TPS, consider any agent $i$ and any $j \not= i$. If $A_j$ contains at least two items then the EFX property (applied two independent times, each time removing a different item from $A_j$) implies that $v_i(A_i) \ge \frac{1}{2} \cdot v_i(A_j)$. Hence if all bundles (except possibly $A_i$) contain at least two items, then $v_i(A_i) \ge \frac{1}{2n-1}v_i(\items) = \frac{n}{2n-1}\cdot Prop(v_i,\items,\frac{1}{n}) \ge \frac{n}{2n-1}\cdot TPS(v_i,\items,\frac{1}{n})$. If there is a bundle $A_j$ that contains only a single item $e$, we may pretend that $e$ is over-proportional (this can only increase the TPS) and remove $e$ and a single agent. The recursive definition of the TPS implies that when we remain in the situation in which every bundle (except possibly $A_i$) has at least two items, the TPS of the remaining instance has not decreased. Hence if $n' < n$ agents remain, we get at least $\frac{n'}{2n'-1}\cdot TPS(v_i,\items,\frac{1}{n}) \ge \frac{n}{2n-1}\cdot TPS(v_i,\items,\frac{1}{n})$. 
\end{proof}

\section{EQX allocations}
\label{app:EQX}



\begin{definition}
    \label{def:EQX}
    An allocation $A_1, \ldots, A_n$ is equitable up to any item (EQX) if for every two agents $i$ and $j$, either $v_i(A_i) \ge  v_j(A_j)$, or for every item $e \in A_j$ it holds that $v_i(A_i) \ge  v_j(A_j \setminus \{e\})$. 
\end{definition}

EQX allocations always exist~\cite{FreemanSVX19}. If all items have identical valuations, then EQX allocations coincide with EFX allocations. This explains Remark~\ref{rem:EFX}.


\section{Randomized protocols for Prop1 allocations}
\label{app:Prop1}

In this section we prove Proposition~\ref{pro:randProp1}.

\begin{proof}
The protocol is a randomized version of the protocol presented in the proof of Proposition~\ref{pro:detProp1}. The difference is that instead of using a deterministic median finding procedure, one uses a randomized one. 
The only randomized aspect of the protocol is that agents are sorted in a uniformly random order $\sigma$. 

Let us recall first a standard randomized algorithm for selecting the median.
Given a list $X$ with $n$ elements in which one seeks the element of order $k$, choose an element $x_i$ uniformly at random to be the {\em splitting element}. Compare every other element to $x_i$, creating two lists $X_h$ (of elements of value higher than $x_i$) and $X_{\ell}$ (of elements of value lower that $x_i$). If $|X_h| = k-1$ then $x_i$ was of order $k$. If $|X_h| \ge k$, continue recursively with $X_h$ instead of $X$, updating $n$ to be $|X_h|$. If $|X_h| < k-1$, continue recursively with $X_{\ell}$ instead of $X$, updating $n$ to be $|X_{\ell}|$ and $k$ to be $k - 1 - |X_h|$.

It is well known that the expected number of splitting elements used by this procedure is at most $O(\log n)$, and the expected number of comparisons is $O(n)$.

Let us now see how the median finding algorithm can be applied in order to find a median cut item. The proposed cut items of the agents serve as the elements. 
Selecting a random element is implemented by selecting a random agent. As the order of $\sigma$ is random, this is done by selecting the lowest rank agent under $\sigma$. The value of the splitting element is announced by the selected agent. This takes $O(\log m)$ bits. Thereafter, each other agent can perform on her own the comparison of the value of her element with that of the splitting element, and announce the result. (It is important to note that ties are broken according to the original names of the agents, not the names under $\sigma$.) This takes one bit per comparison. Hence altogether, the randomized communication complexity for finding the median cut is $O(n + \log n \log m)$.

Recall that the the protocol presented in the proof of Proposition~\ref{pro:detProp1} has $\log n$ phases. Note that the random permutation $\sigma$ has no effect on how agents are split in each phase ($\sigma$ only affects the expected running time, not the outcome). Consequently, by linearity of the expectation operator, the same random permutation $\sigma$ can be used in all phases. 
Using analysis as above, the randomized complexity of phase $t$ is $O(2^{t-1} (n2^{1-t} + \log (n2^{1-t}) \log m))$. Summing over all values of $t$, the randomized complexity is $O(n\log n + n \log m)$.

We remark that a more careful analysis shows that the second term in the randomized complexity can be replaced by $n \log \frac{2m}{n}$, which is significantly smaller than $n \log m$ when $m$ is linear in $n$. However, in any case, the overall upper bound is $O(n\log m)$, due to the term $n \log n$.
\end{proof}

\section{Aprop allocations}
\label{app:Aprop}

In this section we prove Theorem~\ref{thm:Aprop1}.

\begin{proof}
Recall that we may assume that for every agent $i$ it holds that $v_i(\items) = n$ and that for every item $e_j$ it holds that $v_i(e_j) \le 1$.

We describe a protocol for finding Aprop allocations. It has three phases.

We say that item $e_j$ is {\em good} for agent $i$ if $v_i(e_j) \ge \frac{n}{2n-1}$ and moreover, there is some item $e' \not= e_j$ such that $v_i(e_j) + v_i(e') > 1$.  An agent $i$ that receives an item that is good for $i$ gets at least her Aprop value. In the first phase of the protocol we let agents select items that are good for them (if there are such items). Those agents that manage to select a good item can be removed from later phases of the protocol.

Specifically, we scan the agents from $i = 1$ to $n$ and do the following.
If there is a yet unallocated item $e_j$ that is good for $i$, agent $i$ announces its index $j$. In this case, agent $i$ receives $e_j$, and is removed from the remaining parts of the protocol. (In the terminology established before the statement of the theorem, agent $i$ receives a {\em hole}, and an empty {\em block}.) If there is no such good item, agent $i$ 
does not receive any item in this phase.

The communication complexity of the first phase is at most $n \log m$. 

In the second phase, we have a set $\items'$ of items and a set $\agents'$ of agents. $\items'$ contains the items not allocated in the first phase, and $\agents'$ contains those agents that did not receive a good item. Note that we want the final allocation to be Aprop with respect to the original input, and not necessarily with respect to this sub-instance that remains. We say that item $e_j \in \items'$ is {\em large} for agent $i$ if $v_i(e_j) \ge \frac{n}{2n-1}$. Note that for a large item $e_j$ it is not the case that there is some item $e' \not= e_j$ such that $v_i(e_j) + v_i(e') > 1$, as then $e_j$ is good, and it cannot be that agent $i$ reaches the second phase and $e_j \in \items'$. Observe that for every agent $i$, there can be at most one large item (if there are more, {then} all of them are good, not large). 

In the second phase, we create a pool $H$ of holes as follows.
Scan the agents from $i = 1$ to $n$, skipping over agents not in $\agents'$. 
For agent $i$, if it has a large item, and furthermore, this large item $e_j$ is not yet in $H$,  then agent $i$ announces its index $j$. In this case, $e_j$ in moved into $H$ (and creates a hole).  If agent $i$ does not have such an item, 
then no item is moved by $i$ into $H$. No items are allocated in this phase.

The communication complexity of the second phase is at most $n \log m$. 

For the third phase, we have the set $\agents'$ of $n'$ agents and the set $\items'$ of $m'$ items. The set $\items'$ is partitioned into $H$ (generated in the second phase) and $\items' \setminus H$, which we denote by $B$. Importantly, for each agent $i$ there are no good items, there can be at most one large item, and if so, this large item is in $H$ and not in $B$. Within each set $H$ and $B$ separately, the items are ordered in order of increasing index (an order that all agents of $\agents'$ know of).

Each agent $i\in \agents'$ on her own (without communicating with other agents) partitions the set $\items'$ into $n' = |\agents'|$ bundles, where each bundle is an Aprop bundle for $i$. 
For $1 \le j \le n'$, each new bundle $B_j$ in the partition is constructed as follows. If $|H| \ge j$, then the first item in $B_j$ is the $j$th item of $H$. If $|H| < j$, then $B_j$ does not contain any item from $H$. Additional candidate items for $B_j$ come only from $B$, starting with the first item of $B$ that is not yet contained in previous bundles, and continuing in increasing order in $B$. This process stops when $B_j$ becomes an Aprop bundle for $i$. Namely, $v_i(B_j) \ge \frac{n}{2n-1}$, and there is some item $e' {\in} \items \setminus B_j$ {for which} $v_i(B_j) + v_i(e') > 1$. Necessarily, $B_j$ contains at least two items (and at least one of them is from $B$), because $\items'$ does not contain any item that is good for $i$.

If after creating $B'_n$ some items still remain (if so, they must belong to $B$, not to $H$), then they are added to $B'_n$.

We claim that the process that we described indeed creates $n'$ bundles. Observe that $v_i(\items') \ge n'$ (because there are only $n-n'$ items of $\items$ missing from $\items'$, and each such item has value at most~1). We claim that for every $j \le n'$, $v_i(B_j) \le \frac{2n}{2n-1}$. For the sake of contradiction, assume otherwise. Note that the last item $e'$ added to $B_j$ must be from $B$. As there are no large items in $B$, we have that $v_i(e') < \frac{n}{2n-1}$. This implies that  $v_i(B_j \setminus \{e'\}) > \frac{n}{2n-1}$. Hence the bundle $B_j \setminus \{e'\}$ was already Aprop before adding $e'$ to it, which is a contradiction. Our claim implies that the total value of the first $n' - 1$ bundles is at most $(n'-1)\frac{2n}{2n-1} \le (n'-1)\frac{2n'}{2n'-1} = n' - \frac{n'}{2n'-1}$. Hence a value of at least $\frac{n'}{2n'-1} \ge \frac{n}{2n-1}$ remains for $B_{n'}$, making it a $\frac{n}{2n-1}$-TPS bundle. To see that $B_{n'}$ is also a Prop1 bundle, consider two cases. If for all $j < n'$ it holds that $v_i(B_j) \le 1$, then necessarily $v_i(B_{n'}) \ge 1$, making it a Prop1 bundle. Alternatively, if for some $j$ it holds that $v_i(B_j) > 1$, then the last item $e'$ added to $B_j$ has value $v_i(e') > \frac{n-1}{2n-1}$, as otherwise already $B_j \setminus \{e'\}$ is an Aprop bundle. Hence $v_i(B_{n'}) + v_i(e') >  1$, showing that $B_{n'}$ is a Prop1 bundle.

At this stage, each agent $i \in \agents'$ holds a partition of the items into $n'$ Aprop bundles. These partitions have the following property that we refer to as a {\em nesting property}. Consider an agent $i$ with her partition $(B_1, \ldots, B_{n'})$, and an agent $i' \not= i$ with her partition $(B'_1, \ldots, B'_{n'})$. Consider an arbitrary index $k \in \{1, \dots, n'\}$. The nesting property says that  either {$\cup_{j=1}^k B_j \subseteq \cup_{j=1}^k B'_j$ or $\cup_{j=1}^k B'_j \subseteq \cup_{j=1}^k B_j$}. It holds because both sets contain exactly the same items from $H$ (either all of $H$ if $j \ge |H|$, or the first $j$ items of $H$), and each of {them} contains a prefix of the items of $B$. The set that contains the longer prefix contains the other set. 

The fact that the partitions have the nesting property implies that the protocol used in Proposition~\ref{pro:detProp1} to find a Prop1 allocation can be adapted in order to now find an Aprop allocation. We only explain how to adapt phase~1 of the algorithm, as other phases are adapted in a similar way.

Let $j = \lfloor \frac{n'}{2} \rfloor$.
In phase~1, each agent sends a message of length $\log m$, reporting which is the last {item} to enter the bundle $B_j$ of her own partition. Sort these items according to there location in the order in $B$ (if two or more agents reported the same item, break ties in favor of lowered named agent). Let $i$ be the agent whose reported item is in location $j$ in the sorted order, {and} let $e\in B$ be this item. The problem decomposes into two sub-problems. The first sub-problem contains the first $j$ items of $H$, and the items of $B$ up to and including $e$. The second subproblem contains the remaining items of $H$, and the remaining items of $B$. The $j$ agents whose reported item is in location up to $j$ in the sorted order continue with the first subproblem (using only the first $j$ bundles in their partition -- note all items of these bundles are included in the first subproblem), whereas the other agents continue with the second subproblem (using only the last $n' - j$ bundles in their partition -- note all items of these bundles are included in the second subproblem).

The communication complexity of this third phase of our protocol in $n \log m \log n$ (similar to that of Proposition~\ref{pro:detProp1}). Consequently, the overall communication complexity of our protocol is  $(1 + o(1))n \log m \log n$.

To achieve randomized communication complexity of $O(n \log m)$, use in the third phase of our protocol the protocol described in the proof of Proposition~\ref{pro:randProp1}, with adaptations similar to those described when adapting the protocol of Proposition~\ref{pro:detProp1}.
\end{proof}

\section{Randomized protocols for binary valuations}
\label{app:binary}





{In this section, we prove Lemma~\ref{lem:binary}. In our proof we shall make use of the Azuma-Hoeffding inequality for martingales. We state here the version of this inequality that we shall use. 

Let $X_1, \ldots, X_q$ be a sequence of non-negative random variables, each upper bounded by $b$. Suppose further that this sequence is a martingale, in the sense that for every $i$, the expectation $E[X_i]$ of $X_i$ is independent of the realization of $X_1, \ldots, X_{i-1}$. Let $S_q = \sum_{i=1}^q X_i$ denote their sum. Then:


\begin{equation}
    \label{eq:Azuma}
    Pr(S_q - E[S_q] > t) \le e^{-t^2/2qb^2}
\end{equation}
}

\begin{lemma}
\label{lem:binary}
    Consider the value of the parameter $\ell$ in the protocol in the proof of Theorem~\ref{thm:binaryUpper}. When the order $\pi$ of items in $\items$ is uniformly random, the expected value of $\ell$ is at most $4n^{\frac{6}{7}}$. 
\end{lemma}

\begin{proof}
Recall that we use $k$ to denote $\frac{m}{n}$, and that we assume that $k \le n^{\frac{1}{15}}$.
Let $E[\ell]$ denote the expectation of $\ell$, where the expectation is taken over the choice of random order $\pi$.

We first provide non-rigorous analysis explaining why one should expect a bound of $E[\ell] \le O(\sqrt{n})$. Afterwards, we explain what is the main unjustified assumption that is made in the non-rigorous analysis. Then, we provide rigorous analysis, with a somewhat weaker upper bound on $\ell$.

Consider an agent $i$. Let $f_i$ be the first item available for $i$ (not chosen by agents prior to $i$). Let $t_i$ denote the number of items that $i$ selects into its main set. (For simplicity of the presentation, assume that $i \le n-\ell+1$ so that $i$ actually has a main set.) Hence, her main set is $S_i = \{e_{f_i}, \ldots, e_{f_i + t_i - 1}\}$). Recall that $S_i$ must contain $k_i$ of the items that are large for $v_i$. In particular $v_i(e_{f_i + t_i - 1}) = 1$. In total there are $k_i \cdot n$ items that are large for to $v_i$. In any ordering over all items, the average distance between two consecutive large items is at most $\frac{m}{k_i \cdot n}$ (for the first large item, we measure its distance from the beginning of the sequence of items). This seems to indicate that the expected value of $t_i$ (expectation taken over choice of $f_i$) is $\frac{m}{k_i \cdot n} \cdot k_i = \frac{m}{n} = k$. However, this is not true. For example, it might be that $m > 2k_i \cdot n$ and only the last $k_i \cdot n$ items are large, and then the expectation of $t_i$ (over choice of random $f_i$) is $\Omega(m)$. Due to examples like this, we need to assume that the order $\pi$ of items is random. Given this assumption, then for every $f_i$ the expectation of $t_i$ is indeed $\frac{m}{n}$. {(Formally, to have $t_i$ be finite even if $f_i$ is so close to $m$ so that fewer that $k_i$ large items still remain, we use the convention that the sequence $\pi$ is cyclic, with the first item following the last item.)} This time expectation is taken over random permutation of the items, and not over the choice of $f_i$.

Given that each agent is expected to consume $\frac{m}{n}$ items, there are sufficiently many items for all agents, and it appears that $\ell = 0$. However, this is not true. The order of items is random, and the actual value of $t_i$ is a random variable that may deviate from its expectation (be somewhat smaller or larger). Due to these random deviations, there seems to be constant probability that all items suffice (even with some items leftover), and constant probability that they do not suffice. When they do not suffice, we may expect $\ell$ to be of the order of magnitude of at least $\Omega(\sqrt{n})$. This would happen for example if each $t_i$ is $\frac{m}{2n}$ with probability $\frac{1}{2}$ and $\frac{3m}{2n}$ with probability $\frac{1}{2}$. An upper bound of $O(\sqrt{n})$ could follow by providing an upper bound on the variance of each of the $t_i$. Consequently, the intuitive analysis that we presented above suggests that $E[\ell] = O(\sqrt{n})$.

The intuitive analysis that we presented has a major gap. It assumes that $f_i$ is fixed, and a random order $\pi$ of the items is chosen after $f_i$ is fixed. However, in the actual allocation protocol, $\pi$ is chosen first, and it determines $f_i$. For example, if all agents have the same valuation, then $f_i$ will be the first index before which there are $k_i \cdot (i-1)$ large items. If $f_i$ depends on $\pi$ arbitrarily, it is not true that the expectation of $t_i$ is $\frac{m}{n}$. For example, if $k_i = 1$, then {depending on} $\pi$, there are likely to be choices of $f_i$ that force $t_i = \Omega(\frac{m}{n} \cdot \log m)$.

To make our analysis rigorous, we use the fact that $f_i$ does not depend on $\pi$ in a worst case fashion. To see this, we can expose $\pi$ item by item, in a ``need to know" fashion. That is, by the time in which which $f_i$ is determined, we only expose the first $f_i - 1$ items. At this point, the order among the remaining $m - f_i + 1$ items is still random and independent of the choice of $f_i$. To exploit the fact that the ordering of the last $m - f_i + 1$ items is independent of $f_i$, we shall make use of Claim~\ref{claim:binary}. Before presenting this claim, we shall introduce some notation.

{We use the notation $\nu(n)$ to denote a term that decreases at a rate that is faster than any inverse polynomial in $n$. That is, for every $c > 0$, there is a sufficiently large $n_c$ such that for all $n > n_c$ it holds that $\nu(n) < \frac{1}{n^c}$. It is convenient to note that for every constant $d$ it holds that $n^d \cdot \nu(n)$ is still $\nu(n)$. The only effect of multiplying by $n^d$ is a possible increase in the value of $n_c$ referred to above. The use of the notation $\nu(n)$ is justified in our context, as we assume that $n$ is sufficiently large. Thus, if eventually we wish the term $\nu(n)$ to be smaller than say $\frac{1}{n}$, there will be some choice of $n_0$ that will ensure this for all $n > n_0$.}

\begin{claim}
    \label{claim:binary}
    For sufficiently large $n$, for a random permutation $\pi$ over the items, the following holds with probability at least $1 - {\nu(n)}$. For every agent $i$ and for every integer $s$, the number of items that are large for $v_i$ among the last $s$ items is at least $\frac{s}{m} \cdot k_i \cdot n - n^{\frac{4}{7}}$.
\end{claim}

\begin{proof}
    The expected number of large items among the last $s$ items is $\frac{s}{m} \cdot k_i \cdot n$, and the standard deviation is the square root of this, which is significantly smaller than $n^{\frac{4}{7}}$ 
    {(recall that $k_i \le k \le n^{\frac{1}{15}}$, and $n$ is ``sufficiently large")}. Standard bounds on large deviations (that are omitted) show that the probability of a deviation of $n^{\frac{4}{7}}$ is smaller than inverse polynomial in $n$. The claim follows by taking a union bound over $n$ agents times $m$ starting points. Further details of this standard proof are omitted.
\end{proof}

The conclusion of Claim~\ref{claim:binary} fails to hold with probability at most $\nu(n)$. 
Hence, {from now on}, we assume that the conclusion of Claim~\ref{claim:binary} holds. {More generally, we assume that every event that is shown to hold with probability $1 - \nu(n)$ actually holds.}
{Let us clarify the contents of the above assumption. At various points in our proof, the permutation $\pi$ is not uniformly random, but rather uniformly random conditioned on certain good events already happening. Carrying this conditioning explicitly at all parts of the proof is cumbersome. Hence, instead, we shall drop conditioning on good events that happen with probability $1 - \nu(n)$. Due to the fact that we drop this conditioning, we shall need at the end of our proof to add up the probabilities of all bad events for which we dropped the conditioning. As each of them has probability $\nu(n)$ and there will be only polynomially many of them, their total contribution is still $\nu(n)$. If any of these bad events happen, we assume that our proof breaks down completely, and then we only use the bound $\ell \le n-1$. Thus, these low probability bad events contribute only $n \cdot \nu(n) = o(1)$ to $E[\ell]$.}

Recall that for each agent $i$, the number of items that she receives (in the second phase of the protocol) is $t_i$, which is a random variable  whose value depends on the random permutation $\pi$.  For $i \in \{0, \ldots, n\}$ we define the good event $G_i$ as $\sum_{j=1}^i t_j \le m - k \cdot n^{\frac{6}{7}}$.  Fix $q = n - 3n^{\frac{6}{7}}$. We claim that $Pr[G_q] \ge 1 - \nu(n)$. We note that this claim implies Lemma~\ref{lem:binary}, because then $E[\ell] \le (n-q) + o(1) \le 4n^{\frac{6}{7}}$ (where the last inequality holds for sufficiently large $n$), as desired.

It remains to prove the claim that $Pr[G_q] \ge 1 - \nu(n)$. To do so we prove by induction that for every $i \in \{1, \ldots, q\}$, $Pr[G_i] \ge 1 - \nu(n)$. For the base case of the induction we have $Pr[G_0] = 1$, which holds because when no items are selected, then certainly $m - k \cdot n^{\frac{6}{7}}$ still remain.

We now show how to carry out the induction step. We shall do this for the last step of $i = q$, showing that $Pr[G_q] \ge 1 - \nu(n)$. (The proofs for all other steps are the same).


Recall that we assume that Claim~\ref{claim:binary} holds. We may also assume that $G_{q-1}$ holds, as by the inductive hypothesis, this is an event that happens with probability at least $1 - \nu(n)$.


If $G_{q-1}$ holds, then for any agent $i \le q$, $f_i$ is such that the number $s_i = m - f_i +1$ of remaining items is larger than $k \cdot n^{\frac{6}{7}}$, and the number of large items that remain is at least $\frac{s_i}{m} \cdot k_i \cdot n - n^{\frac{4}{7}}$ (by Claim~\ref{claim:binary}). It follows that the expectation of $t_i$ (expectation taken over random order of the remaining items) is at most 

$$s_i \cdot \frac{k_i}{\frac{s_i}{m} \cdot k_i \cdot n - n^{\frac{4}{7}}} \le k \cdot n^{\frac{6}{7}} \cdot \frac{k_i}{\frac{k \cdot n^{\frac{6}{7}}}{k\cdot n} \cdot k_i \cdot n - n^{\frac{4}{7}}} < k(1 + \frac{2}{n^{\frac{2}{7}}})$$ 


We now replace $t_i$ by a random variable $\bar{t}_i$ that is distributed identically to $t_i$, except that if $t_i$ happens to be larger than $k \cdot (\log n)^2$, the value of $\bar{t}_i$ is $k \cdot (\log n)^2$. Three points to notice about $\bar{t}_i$ are the following.

\begin{itemize}
\item $\bar{t}_i$ is a bounded non-negative random variable, of value at most $k \cdot (\log n)^2$.
    \item $E[\bar{t}_i] \le E[t_i] \le k(1 + \frac{2}{n^{\frac{2}{7}}})$.
    \item 
    $Pr[\bar{t}_i \not= t_i] \le e^{-\Omega(\log^2 n)} \le \nu(n)$.
\end{itemize}

Due to the third point above, $Pr[\sum_{i=1}^q t_i \le n - n^{\frac{6}{7}}] \ge Pr[\sum_{i=1}^q \bar{t}_i \le n - n^{\frac{6}{7}}] - \nu(n)$. Hence, to lower bound $Pr[G_q]$ it suffices to lower bound $Pr[\sum_{i=1}^q \bar{t}_i \le n - n^{\frac{6}{7}}]$. A difficulty in the analysis stems from the fact that all $t_i$ (and $\bar{t}_i$) depend on the same permutation $\pi$, and so they are not independent.

To overcome this difficulty, we introduce new random variables $X_1, \ldots, X_q$. Their distributions are selected by an adversary so as to minimize $Pr[\sum_{i=1}^q X_i \le n - n^{\frac{6}{7}}]$. This is done in an online fashion, with the adversary selecting the distribution for $X_i$ only after the values of $X_1, \ldots, X_{i-1}$ are revealed. In selecting the distribution for $X_i$, the adversary needs to satisfy two constraints:

\begin{itemize}
    \item $X_i$ is a bounded non-negative random variable, of value at most $k \cdot (\log n)^2$.
    \item $E[X_i] = k(1 + \frac{2}{n^{\frac{2}{7}}})$.
\end{itemize}

As the adversary may choose $X_i$ to stochastically dominate $\bar{t}_i$, we have that $Pr[\sum_{i=1}^q \bar{t}_i \le n - n^{\frac{6}{7}}] \ge Pr[\sum_{i=1}^q X_i \le n - n^{\frac{6}{7}}]$. Hence it suffices to lower bound $Pr[\sum_{i=1}^q X_i \le n - n^{\frac{6}{7}}]$.

We have that $E[\sum_{i=1}^q X_i] = q\cdot k(1 + \frac{2}{n^{\frac{2}{7}}}) = (n - 3n^{\frac{6}{7}})\cdot k(1 + \frac{2}{n^{\frac{2}{7}}}) \le kn - 2kn^{\frac{6}{7}}$, which is consistent with event $G_q$ occurring, and we are even left with $kn^{\frac{6}{7}}$ items to spare.

To analyse the deviations from expectation of $\sum_{i=1}^q X_i$, we may use Azuma's inequality~\ref{eq:Azuma}, because $X_1, \ldots, X_q$ is a martingale. It implies that the probability of deviating by $k \cdot n^{\frac{6}{7}}$ is at most $e^{\frac{-k^2 \cdot n^{12/7}}{2n \cdot (k \log^2 n)^2}} \le \nu(n)$, completing the proof of Lemma~\ref{lem:binary}. 
\end{proof}









\section{Random bundling}
\label{app:randomBundling}

In this section we present a technique that we refer to as {\em random bundling}. It is applicable only to some fair allocation problems, and when it applies, it can replace dependence on $m$ by dependence on $n$ in the randomized communication complexity. We illustrate this technique in one of the cases in which it is applicable, improving the $O(n\log m)$ upper bound on the randomized communication complexity for $\frac{n}{2n-1}$-TPS to $O(n \log n)$.

\begin{theorem}
    \label{thm:TPSn}
    For additive valuations, the randomized communication complexity for $\frac{n}{2n-1}$-TPS allocations is $O(n \log n)$.
\end{theorem}

\begin{proof}
    Recall that we may assume that for every agent $i$, $v_i(\items) = n$, and that for every item $e$, $v_i(e) \le 1$, and that we seek an allocation that gives each agent value at least $\frac{n}{2n-1}$. We have shown (see Section~\ref{sec:TPS}) that the randomized communication complexity of finding such an allocation is $O(n \log m)$. Moreover, we observe here that the same randomized allocation algorithm works even if the condition $v_i(e) \le 1$ is replaced by the more relaxed condition $v_i(e) \le \frac{2n}{2n-1}$. Armed with this observation, we apply the {\em random bundling} technique.

    Let $m' = \max[2n^9, K]$, where $K$ is some universal constant whose value is determined in the proof of Claim~\ref{claim:randomBundling} below. (With more careful analysis, a lower value of $m'$ also suffices, but this is not needed here.) If $m \le m'$, here is nothing to prove, because $\log m$ is $O(\log n)$. Hence suppose that $m > m'$. In this case, partition the $m$ items into $m'$ bundles uniformly at random, where each item is placed independently 
 uniformly at random in one of the bundles. Let the resulting bundles be $B_1, \ldots, B_{m'}$. We say that this random bundling is {\em good} if for every agent $i$ and bundle $B_j$ it holds that $v_i(B_j) \le \frac{2n}{2n-1}$

    \begin{claim}
    \label{claim:randomBundling}
        With probability at least $\frac{1}{2}$, for every agent $i$ and bundle $B_j$ it holds that $v_i(B_j) \le \frac{2n}{2n-1}$.  
    \end{claim}

    \begin{proof}
        For arbitrary $i$ and $j$, we prove that the probability of the event $v_i(B_j) > \frac{2n}{2n-1}$ is at most $\frac{1}{2nm'}$. The claim then follows by a union bound over all choices of $i$ and $j$.

        An item $e\in \items$ will be referred to as {\em large} if $v_i(e) \ge \frac{1}{n^4}$, and {\em small} otherwise. Let $L$ and $S$ denote the sets of large and small items. Necessarily, $|L| \le n^4$. The probability that $B_j$ contains two or more large items is at most $\binom{|L|}{2} \cdot \frac{1}{(m')^2} < \frac{n^8}{2\cdot 4n^{18}} \le \frac{1}{4nm'}$. 

        Hence, we may assume that $B_j$ contains at most one large item. As no item has value above~1, it suffices to show that the small items contribute total value of at most $\frac{2n}{2n-1} - 1 = \frac{1}{2n-1}$. Observe that their expected contribution is at most $\frac{n}{m'} = \frac{1}{n^8}$. Hence it suffices to show that the probability of deviating from the expectation by my more than $\frac{1}{2n}$ is at most $\frac{1}{4nm'}$. This follows from Hoeffding's inequality. We give more details below.

        Let $X_1, \ldots, X_q$ be independent non-negative random variables, where each $X_i$ is upper bounded by $b_i$. Let $S_q = \sum_{i=1}^q X_i$ denote their sum. Then Hoeffding's inequality states that:

\begin{equation}
    \label{eq:Hoeffding}
    Pr(S_q - E[S_q] > t) \le e^{-2t^2/\sum_i b_i^2}
\end{equation}

For the small items, we have that $b_i \le \frac{1}{n^4}$. As $v_i(\items) = n$, the sum $\sum_i b_i^2$ is maximized if there are $n^5$ small items of value $\frac{1}{n^4}$. Thus $\sum_i b_i^2 \le \frac{1}{n^3}$. As we can take $t = \frac{1}{2n}$, inequality (\ref{eq:Hoeffding}) implies an upper bound of 
$e^{-2t^2/\sum_i b_i^2} \le e^{-n/2}$ on the probability of large deviation. There is some $n_0$ such that for $n \ge n^0$ it holds that $e^{-n/2} \le \frac{1}{4nm'} = \frac{1}{8n^{10}}$, as desired. For $n < n_0$, one takes $m'$ as the universal constant $K = 2n_0^9$.
\end{proof}

We now continue with the description of the protocol. Each agent $i$ is asked to provide one bit of information, specifying whether the random bundling is good for her. This takes $n$ bits of communication. If all agents approve, then treat each bundle as a single item in a new instance with only $m'$ items. Then, an allocation giving each agent value at least $\frac{n}{2n-1}$ can be found with randomized communication complexity of $O(n \log m') \le O(n \log n)$. If not all agents approve, then try a new independent random bundling and repeat. By Claim~\ref{claim:randomBundling}, the expected number of tries is at most~2, and hence, the overall randomized communication complexity is $O(n\log n)$.
\end{proof}

\end{appendix}

\end{document}